\documentclass[longauth]{aa}

\usepackage{graphicx}
\usepackage{amsmath}
\usepackage{amssymb}
\usepackage{comment}
\usepackage{epstopdf}  
\graphicspath{{pdf_figures/}} 
\usepackage{subfig}
\usepackage[usenames,dvipsnames]{xcolor}
\usepackage{multirow}
\usepackage{orcidlink}
\usepackage{txfonts}
\usepackage{hyperref}
\hypersetup{
    colorlinks=true,
    citecolor=blue,
    linkcolor=blue,
    urlcolor=blue, 
    filecolor=blue 
}

\newcommand{\btwo}{{B2\,1811+31}}

\begin{document} 

   \title{Very-high-energy gamma-ray detection and long-term multi-wavelength view of the flaring blazar B2 1811+31}

    \author{\small
    K.~Abe\inst{1} \and
    S.~Abe\inst{2}\orcidlink{0000-0001-7250-3596} \and
    J.~Abhir\inst{3}\orcidlink{0000-0001-8215-4377} \and
    A.~Abhishek\inst{4} \and
    V.~A.~Acciari\inst{5}\orcidlink{0000-0001-8307-2007} \and
    A.~Aguasca-Cabot\inst{6}\orcidlink{0000-0001-8816-4920} \and
    I.~Agudo\inst{7}\orcidlink{0000-0002-3777-6182} \and
    T.~Aniello\inst{8} \and
    S.~Ansoldi\inst{9,42}\orcidlink{0000-0002-5613-7693} \and
    L.~A.~Antonelli\inst{8}\orcidlink{0000-0002-5037-9034}
% ...
    }
    \institute {
    Japanese MAGIC Group: Department of Physics, Tokai University, Hiratsuka, 259-1292 Kanagawa, Japan
    \and Japanese MAGIC Group: Institute for Cosmic Ray Research (ICRR), The University of Tokyo, Kashiwa, 277-8582 Chiba, Japan
    \and ETH Z\"urich, CH-8093 Z\"urich, Switzerland
    \and Universit\`a di Siena and INFN Pisa, I-53100 Siena, Italy
    \and Institut de F\'isica d'Altes Energies (IFAE), The Barcelona Institute of Science and Technology (BIST), E-08193 Bellaterra (Barcelona), Spain
    \and Universitat de Barcelona, ICCUB, IEEC-UB, E-08028 Barcelona, Spain
    \and Instituto de Astrof\'isica de Andaluc\'ia-CSIC, Glorieta de la Astronom\'ia s/n, 18008, Granada, Spain
    \and National Institute for Astrophysics (INAF), I-00136 Rome, Italy
    \and Universit\`a di Udine and INFN Trieste, I-33100 Udine, Italy
    \and Max-Planck-Institut f\"ur Physik, D-85748 Garching, Germany
    \and Universit\`a di Padova and INFN, I-35131 Padova, Italy
    \and Croatian MAGIC Group: University of Zagreb, Faculty of Electrical Engineering and Computing (FER), 10000 Zagreb, Croatia
    \and Centro Brasileiro de Pesquisas F\'isicas (CBPF), 22290-180 URCA, Rio de Janeiro (RJ), Brazil
    \and IPARCOS Institute and EMFTEL Department, Universidad Complutense de Madrid, E-28040 Madrid, Spain
    \and Instituto de Astrof\'isica de Canarias and Dpto. de  Astrof\'isica, Universidad de La Laguna, E-38200, La Laguna, Tenerife, Spain
    \and University of Lodz, Faculty of Physics and Applied Informatics, Department of Astrophysics, 90-236 Lodz, Poland
    \and Centro de Investigaciones Energ\'eticas, Medioambientales y Tecnol\'ogicas, E-28040 Madrid, Spain
    \and Departament de F\'isica, and CERES-IEEC, Universitat Aut\`onoma de Barcelona, E-08193 Bellaterra, Spain
    \and Universit\`a di Pisa and INFN Pisa, I-56126 Pisa, Italy
    \and INFN MAGIC Group: INFN Sezione di Bari and Dipartimento Interateneo di Fisica dell'Universit\`a e del Politecnico di Bari, I-70125 Bari, Italy
    \and Armenian MAGIC Group: A. Alikhanyan National Science Laboratory, 0036 Yerevan, Armenia
    \and Department for Physics and Technology, University of Bergen, Norway
    \and INFN MAGIC Group: INFN Sezione di Torino and Universit\`a degli Studi di Torino, I-10125 Torino, Italy
    \and Croatian MAGIC Group: University of Rijeka, Faculty of Physics, 51000 Rijeka, Croatia
    \and Universit\"at W\"urzburg, D-97074 W\"urzburg, Germany
    \and Technische Universit\"at Dortmund, D-44221 Dortmund, Germany
    \and Japanese MAGIC Group: Physics Program, Graduate School of Advanced Science and Engineering, Hiroshima University, 739-8526 Hiroshima, Japan
    \and Deutsches Elektronen-Synchrotron (DESY), D-15738 Zeuthen, Germany
    \and Armenian MAGIC Group: ICRANet-Armenia, 0019 Yerevan, Armenia
    \and Croatian MAGIC Group: Josip Juraj Strossmayer University of Osijek, Department of Physics, 31000 Osijek, Croatia
    \and Finnish MAGIC Group: Finnish Centre for Astronomy with ESO, Department of Physics and Astronomy, University of Turku, FI-20014 Turku, Finland
    \and University of Geneva, Chemin d'Ecogia 16, CH-1290 Versoix, Switzerland
    \and Saha Institute of Nuclear Physics, A CI of Homi Bhabha National Institute, Kolkata 700064, West Bengal, India
    \and Inst. for Nucl. Research and Nucl. Energy, Bulgarian Academy of Sciences, BG-1784 Sofia, Bulgaria
    \and Japanese MAGIC Group: Department of Physics, Yamagata University, Yamagata 990-8560, Japan
    \and Finnish MAGIC Group: Space Physics and Astronomy Research Unit, University of Oulu, FI-90014 Oulu, Finland
    \and Japanese MAGIC Group: Chiba University, ICEHAP, 263-8522 Chiba, Japan
    \and Japanese MAGIC Group: Institute for Space-Earth Environmental Research and Kobayashi-Maskawa Institute for the Origin of Particles and the Universe, Nagoya University, 464-6801 Nagoya, Japan
    \and Japanese MAGIC Group: Department of Physics, Kyoto University, 606-8502 Kyoto, Japan
    \and INFN MAGIC Group: INFN Roma Tor Vergata, I-00133 Roma, Italy
    \and Japanese MAGIC Group: Department of Physics, Konan University, Kobe, Hyogo 658-8501, Japan
    \and also at International Center for Relativistic Astrophysics (ICRA), Rome, Italy
% ...
}
    
   \date{Received 9 November 2024 / Accepted 20 March 2025}

  \abstract
   {
   Among the blazars whose emission has been detected up to very-high-energy (VHE; $100\,\rm GeV < E < 100\,\rm TeV$) $\gamma$ rays, intermediate synchrotron-peaked BL Lacs (IBLs) are quite rare. The IBL \btwo\ ($z = 0.117$) exhibited intense flaring activity in 2020.
   Detailed characterization of the source emission from radio to $\gamma$-ray energies was achieved with quasi-simultaneous observations, which led to the first-time detection of VHE $\gamma$-ray emission from the source with the MAGIC telescopes.
   }
   {
   In this work, we present a comprehensive multi-wavelength (MWL) view of \btwo, with a specific focus on the 2020 VHE flare, employing data from MAGIC, \textit{Fermi}-LAT, \textit{Swift}-XRT, \textit{Swift}-UVOT and from several optical and radio ground-based telescopes.
   }
   {
   Long-term MWL data were employed to contextualize the high-state episode within the source emissions over 18 years.
   We investigate the variability, cross-correlations and classification of the source emissions during low and high states.
   We propose an interpretative leptonic model for the observed radiative high state.
   }
   {
   During the 2020 flaring state, the synchrotron peak frequency shifted to higher values and reached the limit of the IBL classification.
   Variability in timescales of few hours in the high-energy (HE; $100\,\rm MeV < E < 100\,\rm GeV$) $\gamma$-ray band poses an upper limit of $6 \times 10^{14}\, \delta_\text{D}\, \text{cm}$ to the size of the emission region responsible for the $\gamma$-ray flare, $\delta_\text{D}$ being the relativistic Doppler factor of the region.
   During the 2020 high state, the average spectrum became harder in the HE $\gamma$-ray band compared to the low states. A similar behaviour has been observed in X rays.
   Conversely, during different activity periods, we find harder-when-brighter trends in X rays and a hint of softer-when-brighter trends at HE $\gamma$ rays.
   Long-term HE $\gamma$-ray and optical correlation indicates the same emission regions dominate the radiative output in both ranges, whereas the evolution at 15 GHz shows no correlation with the fluxes at higher frequencies.
   We test one-zone and two-zone synchrotron-self-Compton models for describing the broad-band spectral energy distribution during the 2020 flaring state and investigate the self-consistency of the proposed scenario.
   }
   {}


   \maketitle

\nolinenumbers

\section{Introduction}
\label{sec:intro}

Blazars are active galactic nuclei (AGNs) with their relativistic jets pointing towards the observer. Their central engine is likely to be a supermassive black hole of $10^{7} - 10^{9}$ solar masses, fed by the infall of matter from a surrounding accretion disk. 
Although constituting a tiny fraction among all astrophysical sources observed in the optical band, blazars are by far the most common type of objects detected at $\gamma$-ray energies \citep{2020ApJS..247...33A}.

Radiative emission from blazars is mostly non-thermal radiation and it ranges from radio to very-high-energy (VHE; $100\,\rm GeV < E < 100\,\rm TeV$) $\gamma$ rays. 
The broad-band spectral energy distribution (SED) of blazars is characterized by two distinct bumps \citep[e.g.][]{Ghisellini2017}. 
The low-energy one peaks in the infrared-to-X-ray energy range and it is commonly attributed to synchrotron radiation emitted by relativistic electrons accelerated in the jet.
The high-energy bump peaks above MeV energies and it is most likely due to inverse Compton (IC) scattering.
The photon seeds for the IC scattering can be the synchrotron ones from the same electron population \citep[synchrotron self-Compton, SSC;][]{1981ApJ...243..700K, 1992ApJ...397L...5M} or can be external to the jet, such as radiation from the broad-line region (BLR), accretion disk and dusty molecular torus \citep{1973A&A....24..337S, 1994ApJS...90..945D, 2016ApJ...830...94F}.
The presence of a sub-dominant hadronic component is also possible, as discussed in \cite{2000NewA....5..377A} and \cite{Murase_2012} and suggested by the evidence for neutrino emission from active galaxies \citep{2018Sci...361.1378I, 2018Sci...361..147I, 2018ApJ...863L..10A, 2022Sci...378..538I}.

According to the features in their optical/UV spectra, blazars can be classified as flat spectrum radio quasars (FSRQs), which are characterized by the presence of strong emission lines in their optical/UV spectra, and BL Lacertae objects (BL Lacs), having no or weak emission lines.
BL Lacs can be further divided into three sub-classes based on their synchrotron peak frequency $\nu_{\text{s}}$, i.e. low-frequency-peaked (LBLs, $\nu_{\text{s}} < 10^{14}\,\text{Hz}$), intermediate-frequency-peaked (IBLs, $10^{14}\,\text{Hz}<\nu_{\text{s}}<10^{15}\,\text{Hz}$), and high-frequency-peaked (HBLs, $\nu_{\text{s}} > 10^{15}\,\text{Hz}$) BL Lacs \citep[e.g.][]{1995ApJ...444..567P}.
Most of the blazars detected up to VHE $\gamma$ rays are HBLs.
At the time of writing (October 2024), according to TeVCat\footnote{\url{http://tevcat.uchicago.edu/}}, VHE $\gamma$-ray emission has been detected from 57 HBLs.
Conversely, only 10 IBLs and 2 LBLs have been detected at VHE $\gamma$ rays, usually during flaring episodes \citep[e.g.][]{2018A&A...619A..45M}.
In high-energy (HE; $100\,\rm MeV < E < 100\,\rm GeV$) $\gamma$ rays, the majority of the BL Lac objects detected by the Large Area Telescope (LAT) onboard the \textit{Fermi Gamma-Ray Space Telescope} are LBLs and IBLs \citep[e.g.][]{2022ApJS..263...24A}.
The lack of low-frequency peaking sources in the VHE $\gamma$-ray band is mainly due to their high-energy bump being located at lower energies than for HBLs.

Blazars are characterized by high variability over very different timescales.
Light-curves can show long-term trends on periods lasting from several years to months and short-term activity on timescales from weeks to days. 
For several blazars, intra-night/day variability on hours-to-minutes timescales has been assessed \citep[e.g.][]{1988ApJ...325..628S, 2018ApJ...863..175G}.
The fastest variations are observed especially during flaring episodes.
The amplitudes of the variations in time of the radiative emissions of blazars are energy-dependent.
In the case of the HBL Mrk\,421, the largest variability amplitudes have been detected in the energy bands corresponding to the falling tails of the two bumps of the SED, in hard X rays (from few keV up to several tens of keV) and in the VHE $\gamma$-ray band \citep[e.g.][]{2020ApJS..248...29A}.

In order to perform detailed studies of the radiative emission mechanisms acting in blazars, it is fundamental to have a complete energy coverage of their emission, with multi-wavelength (MWL) observations covering from radio to VHE.
As blazars may show extremely fast variable behavior during flares, it is of crucial importance that these observations be performed (quasi-) simultaneously, in order to provide a proper characterization of the source emission state. However, especially for weak blazars such as the one studied in this work, complete coverage may be difficult to obtain in all energy ranges during low emission states.

The blazar \btwo, located at R.A. and Dec. (J2000) 18h13m35.2028s, +31d44m17.621s \citep{2011AJ....142..105P}, is classified as a BL Lac object in the \textit{Fermi}-LAT Fourth Source Catalog \citep[4FGL, ][]{2020ApJS..247...33A} and as an IBL in \cite{1999ApJ...525..127L}.
The blazar was listed as one of the most promising VHE $\gamma$-ray candidates in \cite{2017A&A...608A..68F}. Optical spectroscopic observations reported in \cite{1991ApJ...378...77G} settled its redshift to $z = 0.117$.

Following the detection by the \textit{Fermi}-LAT of a high state from the source in the $E > 100\,\text{MeV}$ energy range on October 1, 2020 (MJD 59123) \citep[][]{2020ATel14060....1A}, a MWL observational campaign on \btwo\ was organized.
The observations performed during this high-state period with the Major Atmospheric Gamma-ray Imaging Cherenkov (MAGIC) telescopes led to the first-time detection of VHE $\gamma$-ray emission from the source \citep{2020ATel14090....1B}.
The telescopes on board the \textit{Neil Gehrels Swift Observatory}, which are sensitive in the optical-to-X-ray range, joined the follow-up campaign, as well as several optical and radio ground-based telescopes.
These observations allowed us to characterize in detail the properties of the source high state from radio to VHE $\gamma$ rays.

In addition, we include in this paper a comprehensive MWL dataset from 2005 up to 2024 extending from radio to HE $\gamma$ rays.
\btwo\ was a frequent target of \textit{Swift} observations since 2005. It has been monitored since 2009 at 15 GHz by the Owens Valley Radio Observatory (OVRO) and it was included over the years in several monitoring programs in the optical band.
Ultimately, \btwo\ has been observed by \textit{Fermi}-LAT for its entire mission thanks to the continuous sky-survey operations. The source has been significantly detected in HE $\gamma$ rays already in its first months of operations and was already included in the \textit{Fermi}-LAT First Source Catalog \citep[1FGL,][]{2010ApJS..188..405A}.
We present the analyses of these long-term MWL data to contextualize the high-state episode and to compare the spectral and temporal features of the flaring and steady low-state emissions.

The paper is structured as follows.
Section \ref{sec:analysis} is dedicated to present the MWL dataset. The analysis results of data from MAGIC, \textit{Fermi}-LAT, \textit{Swift}-XRT and \textit{Swift}-UVOT are presented, along with the reduction of the optical and radio data.
Section \ref{sec:variability_cross-correlations} is dedicated to the variability, intra-band and multi-band correlation analyses carried out to infer insights about the emission regions responsible for the MWL emission.  
In Section \ref{sec:source_classification}, we report on the classification of the source high and low states according to the partition based on the synchrotron peak frequency. 
A discussion on the literature on the source redshift, along with the redshift indirect estimation from simultaneous HE and VHE observations, is presented in Section \ref{sec:redshift}.
In Section \ref{sec:modeling}, a leptonic interpretation model for the SED reconstructed in quasi-simultaneous observations is presented and its self-consistency and physical implications are discussed.
We summarize the main results and present our conclusions in Section \ref{sec:conclusions}.

\section{Instruments and analysis}
\label{sec:analysis}

In this section we present the MWL datasets and analyses performed in each energy band of \btwo.
Table \ref{summ_table_b2} reports the list of the instruments whose data from the 2020 $\gamma$-ray high state were included in the analysis, along with the time ranges of the observations.
From the same instruments, we collected the available long-term data.
In addition, we included long-term optical data from the Katzman Automatic Imaging Telescope \cite[KAIT,][]{2001ASPC..246..121F}, the Catalina Real-Time Transient Survey \cite[CRTS,][]{2009ApJ...696..870D} and from the Tuorla blazar monitoring program\footnote{\url{https://users.utu.fi/kani/1m/}} \citep{nilsson2018}.
Further details on each dataset are given in the following sections.

\begin{figure*}
\centering
\begin{tabular}{c}
\includegraphics[width = 0.98\textwidth]{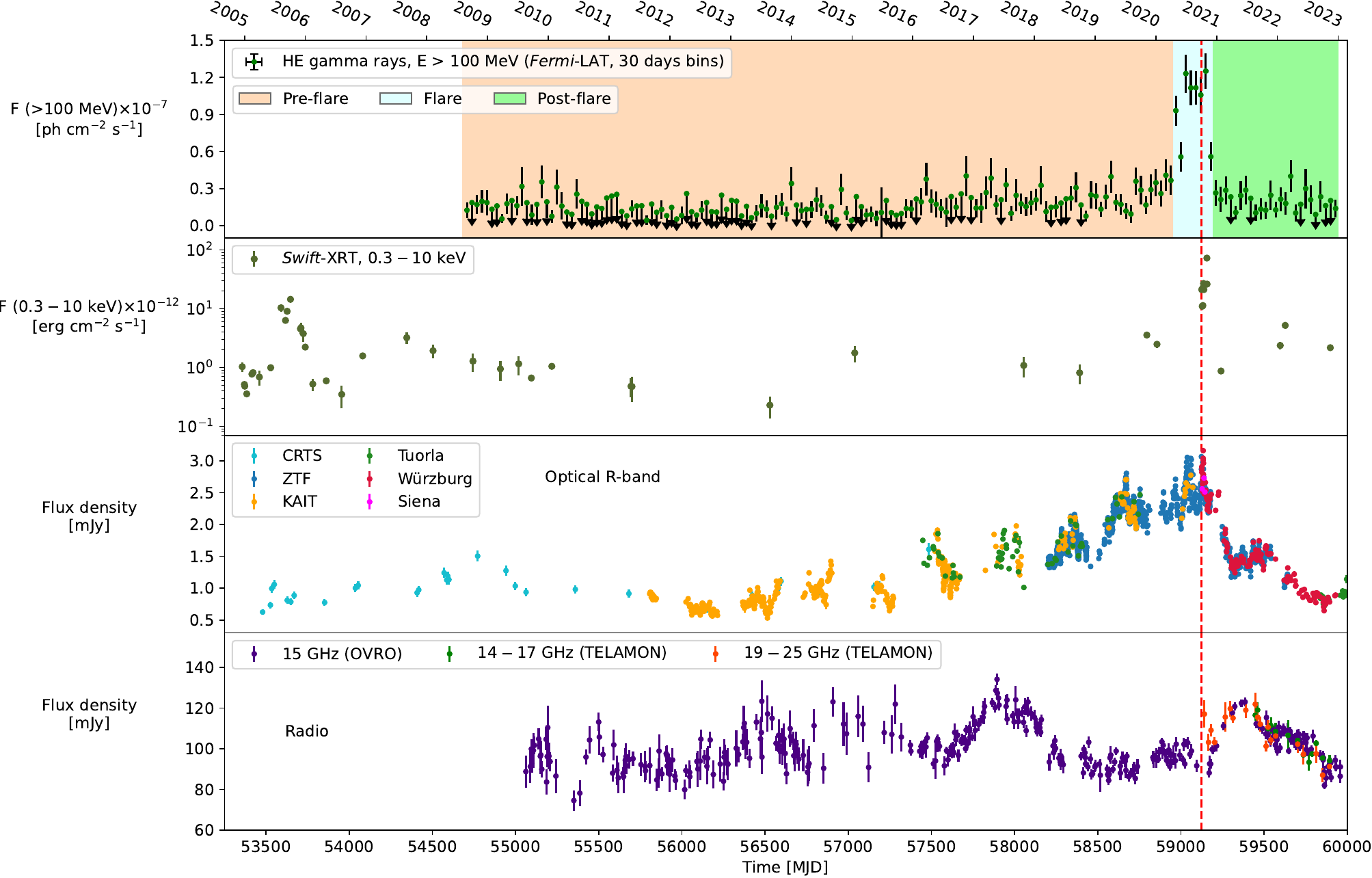}
\end{tabular}
\caption{The \btwo\ long-term light-curve collected from 2005 to 2023. \textit{From top to bottom panels}: HE $\gamma$-ray flux above 100 MeV from \textit{Fermi}-LAT monthly-binned data, X-ray flux in the $0.3-10\,$keV energy range from \textit{Swift}-XRT, optical R-band data and radio data. The red dashed line marks the \textit{Fermi}-LAT high-state detection on October 1, 2020 (MJD 59123).
The light orange, light blue and green shaded bands in the top panel indicate the `Pre-flare', `Flare' and `Post-flare' periods, respectively (Table \ref{tab:summ_table_spectral_B2}).
}
\label{fig:B2_long_term_LC}
\end{figure*}

\begin{figure*} [t]
\centering
\begin{tabular}{c}
\includegraphics[width = 0.92\textwidth]{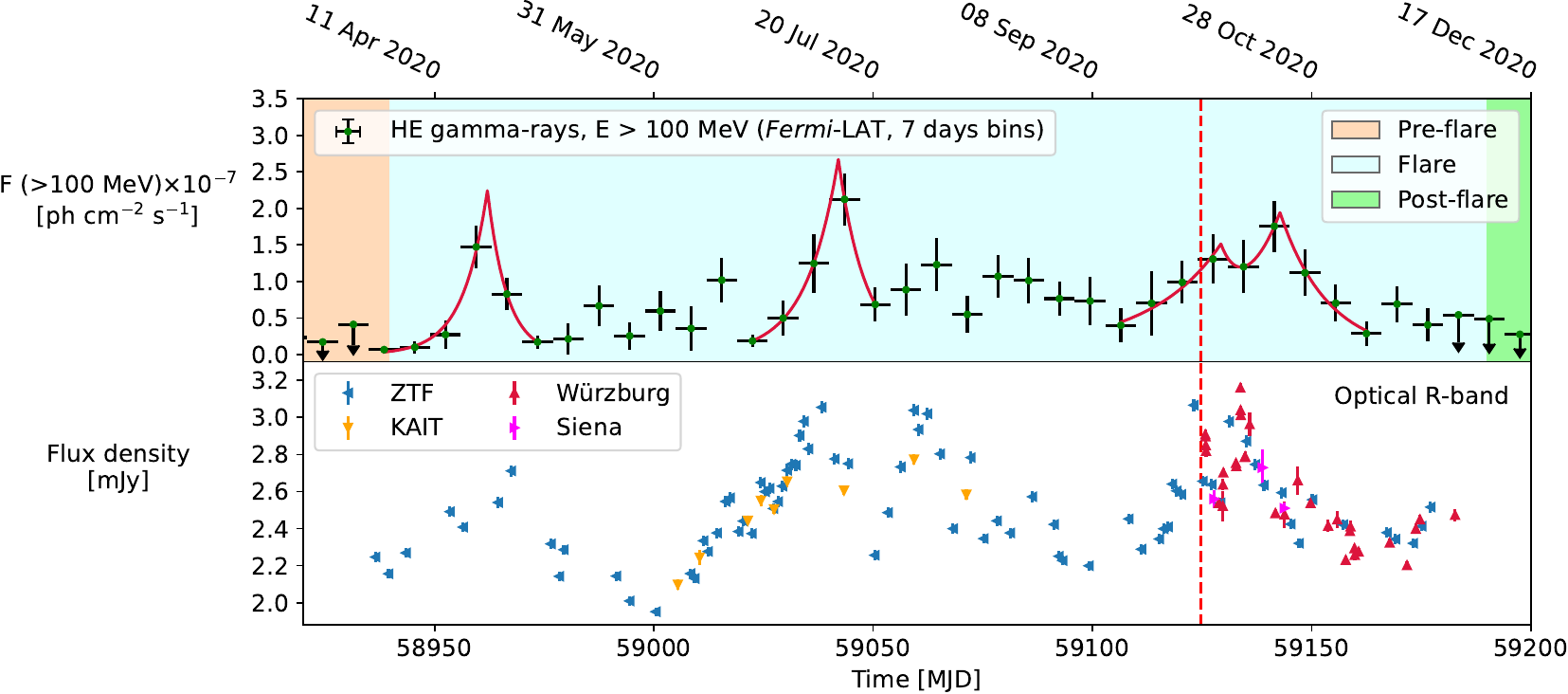}
\end{tabular}
\caption{MWL light-curve of \btwo\ in HE $\gamma$-rays (weekly-binned \textit{Fermi}-LAT data, \textit{upper panel}) and optical R-band (instruments are reported in the legend, \textit{lower panel}) in a period of around 250 days surrounding the October 2020 high state. This period corresponds approximately to the ‘Flare’ period (Table \ref{tab:summ_table_spectral_B2}).
In the upper panel, the solid red lines show the double-exponential fit (Eq. \ref{eq:double_exponential}) of the rising and falling trends in the \textit{Fermi}-LAT light-curve (Section \ref{subsec:variability_weekly}).}
\label{fig:b2_LAT_weekly_optical}
\end{figure*}

\begin{figure*}
\centering
\begin{tabular}{c}
\includegraphics[width=0.95\textwidth]{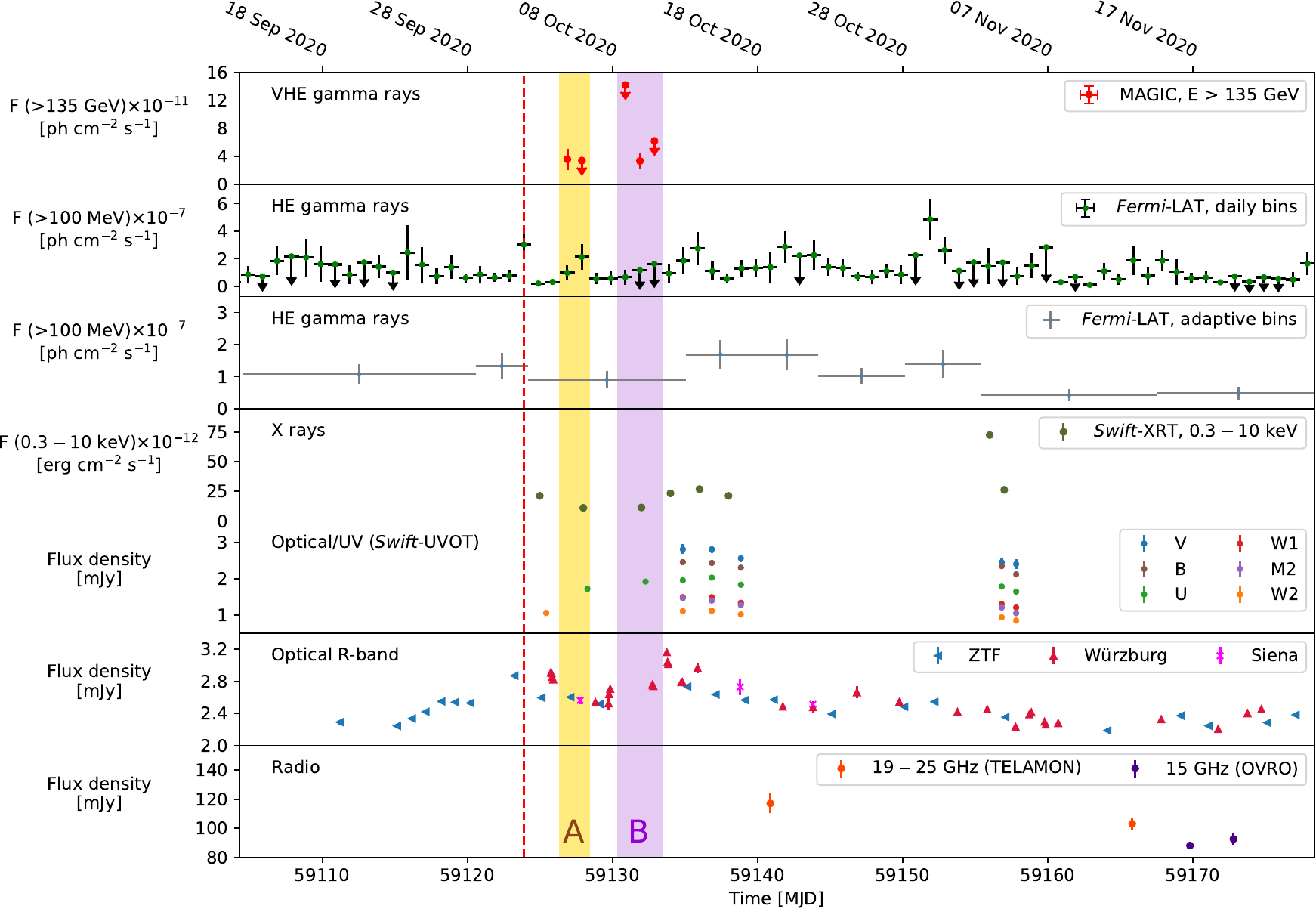}
\end{tabular}
\caption{MWL light-curve of \btwo\ in a period of approximately 70 days surrounding the $\textit{Fermi}$-LAT high-state detection on MJD 59123, marked by the red dashed line.
\textit{From top to bottom panels}: 
VHE $\gamma$-ray flux above 135 GeV from MAGIC, 
HE $\gamma$-ray flux above 100 MeV from \textit{Fermi}-LAT in daily bins and using the `adaptive-binning' method, 
X-ray flux in the $0.3-10\,$keV range from \textit{Swift}-XRT, optical/UV data in the photometric filters of \textit{Swift}-UVOT, optical R-band data and radio data. The yellow and light violet shaded bands indicate periods of 48 h and 72 h surrounding the two sets of MAGIC consecutive observation nights, respectively defined as Periods A and B in Section \ref{sec:modeling}.
}
\label{fig:B2_flare_LC}
\end{figure*}

The long-term MWL light-curve of \btwo\ collected from 2005 to 2023 is shown in Figure \ref{fig:B2_long_term_LC}. 
Figure \ref{fig:b2_LAT_weekly_optical} shows a close-up view of the $\textit{Fermi}$-LAT and optical R-band light-curves over a 250-day period which approximately corresponds to the source HE $\gamma$-ray high state in 2020 (Section \ref{subsec:Fermi_analysis}). 
Additionally, Figure \ref{fig:B2_flare_LC} provides a zoom-in on the MWL light-curve over a 70-day period covering the MAGIC observations and the MWL observational campaign following the high-state detection by $\textit{Fermi}$-LAT on October 1, 2020 (MJD 59123).

\begin{table}
\tiny
\caption{
Instruments participating to the MWL campaign on \btwo\ during the 2020 $\gamma$-ray flare.
}
\label{summ_table_b2}
\begin{center}
\begin{tabular}{ccccc}
\hline
\multicolumn{1}{c}{Instrument}   &
\multicolumn{1}{c}{Sensitivity range} &
\multicolumn{1}{c}{MJD start}    &
\multicolumn{1}{c}{MJD stop}     \\
\hline
MAGIC & $20\,\text{GeV}-100\,\text{TeV}$ & 59127 & 59133 \\
\textit{Fermi}-LAT & $30\,\text{MeV}-1\,\text{TeV}$ & 58940 & 59190 \\ 
\textit{Swift}-XRT & $0.2-10\,\text{keV}$ & 59125 & 59158 \\
\textit{Swift}-UVOT & \textit{v, b, u, w1, m2, w2} & 59125 & 59158 \\
& ($\approx190-600\,\text{nm}$) & & \\
ZTF & \textit{g, r, i} & 59111 & 59180 \\  
& $(\approx 400-800\,\text{nm})$ &  &  \\ 
W{\"u}rzburg, Siena & R & 59126 & 59174 \\
& $(\approx520-800\,\text{nm})$ & & \\ 
TELAMON & $14-25\,\text{GHz}$ & 59141 & 59166 \\
OVRO & $15\,\text{GHz}$ & 59169 & 59177 \\
\hline
\end{tabular}
\tablefoot{
For the optical and UV telescopes, the photometric filters employed in this study are reported.
\textit{Fermi}-LAT detected the high state on October 1, 2020 (MJD 59123). 
Since LAT operates in sky survey mode, we report MJD start and stop times corresponding approximately to the beginning and end of the 2020 $\gamma$-ray flare (Figure \ref{fig:B2_long_term_LC}).
For all the other instruments, the MJD start and stop refer to the observations carried out in the MJD $59100 - 59180$ period (Figure \ref{fig:B2_flare_LC}).
}
\end{center}
\end{table}  

\subsection{MAGIC}
\label{subsec:MAGIC_analysis}
The MAGIC telescopes \citep{Aleksi2016} constitute a stereoscopic system of two 17 m diameter imaging atmospheric Cherenkov telescopes located at the height of about 2200 m a.s.l. at the Observatorio del Roque de los Muchachos (La Palma, Spain).
Following the high-state detection from \textit{Fermi}-LAT in the HE $\gamma$-ray band, the MAGIC telescopes performed observations of \btwo\ from October 5, 2020 (MJD 59127), up to and including October 11, 2020 (MJD 59133). A total of about 5 hours of good quality data in good weather conditions, dark time and wide zenith range from 20$^{\circ}$ up to 65$^{\circ}$ was collected over five observation nights.
Table \ref{tab:magic_table} reports the night-wise time intervals and zenith ranges of the MAGIC observations.
Data were analyzed using the MAGIC analysis and reconstruction software MARS \citep{Zanin2013}. 
We used the standard variable $\theta^2$, which is defined as the squared angular distance of the reconstructed shower direction with respect to the source direction, to look for any significant VHE $\gamma$-ray excess with respect to the background.
Primarily, a unique dataset was obtained by combining the data collected during the 5 observation nights.
On this dataset, the presence of VHE $\gamma$-ray emission from \btwo\ was established with a statistical significance of $5.3\sigma$.
The statistical significance was estimated using the Li\&Ma formula reported in \cite{1983ApJ...272..317L}.

We then derived the night-wise VHE $\gamma$-ray flux for energies above $135$ GeV. 
The energy threshold was optimized in order to achieve a proper flux estimation for each night regardless of the observational conditions, such as zenith range, weather, night sky background level.
The MAGIC light-curve at $E>135\,\text{GeV}$ is reported in Table \ref{tab:magic_table}, along with the night-wise significances of the VHE $\gamma$-ray signal from \btwo. The light-curve is shown in the top panel of Figure \ref{fig:B2_flare_LC}, where 95\% confidence level upper limits are indicated as downward arrows in VHE $\gamma$ rays when the significance is below 3$\sigma$.
Since the flux levels in each observation nights are compatible within the $1\sigma$ statistical uncertainty band, we conclude that no significant variability is seen in the VHE light-curve.
The weakness of the signal prevented any further quest for intra-night variability. 
For reasons that are discussed in Section \ref{subsec:identification_of_compatible_source_states}, the MAGIC dataset was divided into two periods, one including the observations carried out on October 5 and October 6 (Period A), and the second one from October 9 up to and including October 11 (Period B).
For each period, we evaluated the overall spectrum combining the data taken in the observations within the same period and fitted it with a power-law (PL) function
\begin{equation}
    \frac{dN}{dE} = N_{0} \left( \frac{E}{E_{0}} \right)^{- \Gamma_{\text{PL}}}\,,
    \label{eq:PL}
\end{equation}
with photon index $\Gamma_{\text{PL}}$, normalization constant $N_{0}$ and decorrelation energy $E_{0}$.
In order to reconstruct the intrinsic spectra of the source in the two periods, the observed spectra were unfolded by the energy dispersion using the Tikhonov method \citep{2007NIMPA.583..494A} and then corrected for $\gamma$-ray absorption by the interaction with the extra-galactic background light (EBL) using the \cite{dominguez_ebl} model.
The intrinsic spectra are soft with $\Gamma_{\text{PL}} = 4.16 \pm 0.63_\textrm{stat}$, $E_0 =  130.95\, \textrm{GeV}$ and $N_0 = (4.21 \pm 1.51_\textrm{stat})\times 10^{-10}\, \textrm{TeV}^{-1} \, \textrm{cm}^{-2} \, \textrm{s}^{-1}$ for Period A and $\Gamma_{\text{PL}} = 3.75 \pm 0.40_\textrm{stat}$, $E_0 =  125.16\, \textrm{GeV}$ and $N_0 = (7.36  \pm 1.99_\textrm{stat})\times 10^{-10}\, \textrm{TeV}^{-1} \, \textrm{cm}^{-2} \, \textrm{s}^{-1}$ for Period B (Table \ref{tab:summ_table_spectral_MAGIC_B2_stateA_B}).
The soft spectra in the VHE $\gamma$-ray range suggests that the HE bump of the SED is likely to be peaking at lower energies.

\subsection{\textit{Fermi}-LAT}
\label{subsec:Fermi_analysis}
The Large Area Telescope (LAT) on board the \textit{Fermi} satellite is an imaging, wide field-of-view ($\sim2.4\text{sr}$ at few GeV), pair conversion $\gamma$-ray instrument sensitive to photons from $30$ MeV to about $1$ TeV \citep{2009ApJ...697.1071A}.
The \textit{Fermi}-LAT observed a hard-spectrum GeV flare from the source on October 1, 2020 (MJD 59123). The daily $\gamma$-ray flux at $E > 100\,$MeV was found to be higher by a factor 11 than the average value reported in the 4FGL, with a daily photon index of $1.4\,\pm\,0.2$, harder than the $2.14\,\pm\,0.6$ value reported in the 4FGL. In addition, \textit{Fermi}-LAT detected several $E>10\,$GeV photons, the most energetic one having energy of $61\,$GeV, with probability $>99\%$ of being emitted from \btwo\,\citep[][]{2020ATel14060....1A}.

The analysis of \textit{Fermi}-LAT data was performed using the \textit{ScienceTools} v2.0.8 and the {\fontfamily{cmtt}\selectfont fermipy}\footnote{\url{https://fermipy.readthedocs.io/en/latest/} \label{url:SSDC}} v1.0.1 Python package.
Photon events from the Pass 8 P8R3 \citep{2013arXiv1303.3514A, P8R3_ref2} SOURCE class, with reconstructed energy in the $100\,\text{MeV} - 1\,\text{TeV}$ range and direction within 15$^{\circ}$ from the nominal position of the source were selected, applying standard quality cuts ({\fontfamily{cmtt}\selectfont `DATA\_QUAL>0 \&\& LAT\_CONFIG==1'}). The P8R3\_SOURCE\_V3\ instrument response functions were adopted.
To reduce the contamination from the Earth limb, a zenith angle cut of 90$^{\circ}$ was used.
All the localized sources included in the Third Data Release of the 4FGL \citep[4FGL-DR3,][]{2022ApJS..260...53A} with position within 20$^{\circ}$ from the source were included in the model, along with the Galactic interstellar diffuse and residual background isotropic emission, as modeled respectively in gll\_iem\_v07.fits and iso\_P8R3\_SOURCE\_V3\_v1.txt\footnote{\url{https://fermi.gsfc.nasa.gov/ssc/data/access/lat/BackgroundModels.html}}.

We evaluated the significance of the $\gamma$-ray signal from the source employing a maximum-likelihood test statistic (TS) defined as $\text{TS} = 2 \left( \log L_1 - \log L_0 \right)$, where $L$ is the likelihood of the data given the model with ($L_1$) or without ($L_0$) a point-like source at the position of \btwo.

\begin{figure}
\centering
\begin{tabular}{c}
\includegraphics[width=0.95\columnwidth]{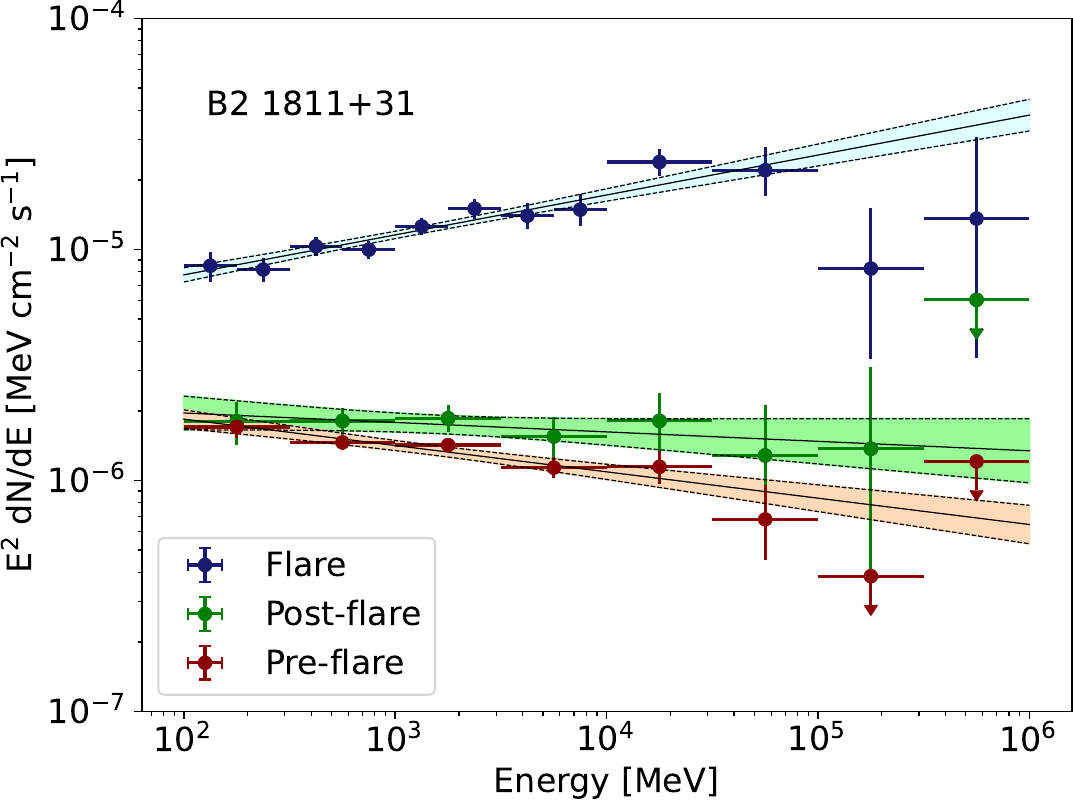}
\end{tabular}
\caption{SED in the $100\,\text{MeV}-1\,\text{TeV}$ energy range resulting from $\textit{Fermi}$-LAT data from the `Pre-flare' (red), `Flare' (blue) and `Post-flare' (green) periods, as defined in Table \ref{tab:summ_table_spectral_B2}.
The average spectral indices and fluxes in the same energy range for the three periods are reported in Table \ref{tab:summ_table_spectral_B2}.}
\label{fig:LAT_preflare_flare_postflare}
\end{figure}

Dedicated spectral analyses were performed in time windows corresponding to the low and high states of the source.
Table \ref{tab:summ_table_spectral_B2} provides the definition of the `Flare' period, corresponding to the 2020 $\gamma$-ray high state, as well as the definitions of the `Pre-flare' and `Post-flare' time intervals.
The `Pre-flare' period lasts from August 2008, i.e. the beginning of \textit{Fermi}-LAT data taking, up to the beginning of the 2020 outburst.
The `Post-flare' extends from the end of the 'Flare' period up to January 2023.
The temporal boundaries of these time intervals were evaluated from the long-term \textit{Fermi}-LAT light-curve as discussed later in this section.
In the top panel of Figure \ref{fig:B2_long_term_LC}, the three periods are indicated by the shaded areas.
In each of these time intervals, a binned maximum-likelihood analysis was performed dividing the energy range in 8 bins per decade and the angular coordinates' space in pixels with size of 0.1$^{\circ}$. In the fit, the normalizations of the diffuse components and of the significantly detected (TS > 25) sources within 5$^{\circ}$ from \btwo\ were left free.
The spectral parameters of the other 4FGL sources were fixed to the published 4FGL-DR3 values.
The HE $\gamma$-ray spectrum of the source was primarily modeled with a PL function (Eq. \ref{eq:PL}) with free normalization and spectral index. The resulting SEDs in the `Pre-flare', `Flare' and`Post-flare' periods are shown in Figure \ref{fig:LAT_preflare_flare_postflare}, whereas the PL best-fit parameters are reported in Table \ref{tab:summ_table_spectral_B2}.

\begin{table}
\small
\caption{\textit{Fermi}-LAT spectral analyses on \btwo\ during the `Pre-flare', `Flare' and `Post-flare' periods.}
\label{tab:summ_table_spectral_B2}
\begin{center}
\begin{tabular}{ccccccc}
\hline
\multicolumn{1}{c}{Period} &
\multicolumn{1}{c}{Start} &
\multicolumn{1}{c}{Stop} &
\multicolumn{1}{c}{$\Gamma_{\text{PL}}$} &
\multicolumn{1}{c}{F (>$100\,\text{MeV}$)$\times$10$^{-8}$ }\\
\multicolumn{1}{c}{}&\multicolumn{1}{c}{[MJD]}&\multicolumn{1}{c}{[MJD]} & \multicolumn{1}{c}{} & \multicolumn{1}{c}{[ph cm$^{-2}$ s$^{-1}$]} \\
\hline
Pre-flare & 54682 & 58940 & $2.11\pm0.03$ & $1.7\pm0.1$\\
Flare & 58940 & 59190 & $1.83\pm0.02$ & $9.6\pm0.5$\\
Post-flare & 59190 & 59945 & $2.04\pm0.05$ & $1.9\pm0.3$\\
\hline
\end{tabular}
\end{center}
\end{table}

During the high state, on average the source experienced strong flux enhancement in the HE $\gamma$-ray band, by a factor of $\sim6$ compared to the average radiative states before and after the 2020 flare. In addition, strong spectral hardening occurs, as the photon index changes from $2.11\pm0.03$ in low state to $1.83\pm0.02$ in high state. These evidence indicate that the spectral break of the HE bump shifted to higher energies with respect to the quiescent state. We provide a qualitative interpretation of this evidence in Section \ref{sec:source_classification}, in view of the classification of the source broad-band emission during the high and low states.

To test the curvature of the HE $\gamma$-ray spectra during the three periods, maximum-likelihood fits were performed assuming log-parabola (LP), power-law with exponential cutoff (PLEC) and broken power-law (BPL) models\footnote{For the LP, PLEC and BPL models, we used the implementations of the {\fontfamily{cmtt}\selectfont LogParabola}, {\fontfamily{cmtt}\selectfont PLSuperExpCutoff} and {\fontfamily{cmtt}\selectfont BrokenPowerLaw}, respectively, as defined in \url{https://fermi.gsfc.nasa.gov/ssc/data/analysis/scitools/source_models.html}}.
The $\text{TS}_{\text{curv}}$ test statistic, defined as $\text{TS}_{\text{curv}} = 2 \left( \log L_{\text{curved model}} - \log L_{\text{PL}} \right)$ as in 4FGL, was employed.
Results are summarized in Table \ref{tab:summ_table_spectral_B2_LP_PLEC_BPL}. No significant improvement was found up to the $3\sigma$ significance level\footnote{
$\text{TS}_{\text{curv},\text{ i}} = 2 \left( \log L_{\text{i}} - \log L_{\text{PL}} \right) < 9$, for $\text{i} = \text{LP,\, PLEC,\, BPL}$, in the `Pre-flare', `Flare' and `Post-flare' periods (Table \ref{tab:summ_table_spectral_B2})}.
The spectral break of the HE bump of 2020 flare SED is reasonably lying around tens of GeV, where the uncertainties become relevant given the $\textit{Fermi}$-LAT sensitivity in this energy range.
Nevertheless, the MAGIC data presented in Section \ref{subsec:MAGIC_analysis} are able to characterize the SED above the spectral break, therefore we postpone a detailed investigation of the break point of the HE SED bump to Section \ref{sec:modeling}, where the modeling of the high-state broad-band emission is presented.

To investigate the long-term variability of \btwo\ in the HE $\gamma$-ray band, the light-curve from August 2008 to January 2023 was evaluated in bins of 30 days by performing dedicated maximum-likelihood fits in each time bin. 
In each time bin, the fit was performed leaving free the normalization and photon index of the \btwo\ PL spectral model, whereas all the spectral parameters of the other sources included in the model were fixed to the best-fit values obtained from the analysis of \textit{Fermi}-LAT data integrated from August 2008 to January 2023.
The light-curve is shown in Figure \ref{fig:B2_long_term_LC}. Upper limits at 95\% confidence level are shown as downward arrows for each time bin in which the $\text{TS}$ of the source was found to be smaller than 4, which corresponds approximately to the $2\sigma$ significance level. As a sanity check, we verified that the flux to flux error ratio is approximately proportional to the $\sqrt{TS}$ for all time bins, as prescribed in \cite{2023ApJS..265...31A}. We repeated the same sanity check also for the other light-curves derived from \textit{Fermi}-LAT data throughout this work.
The long-term light-curve shows that the source experienced a high-state period starting several months before the \textit{Fermi}-LAT high-state detection on MJD 59123.
In Table \ref{tab:summ_table_spectral_B2}, we report the beginning of the `Flare' period to be MJD 58940. This date approximately corresponds to the first bin of the monthly \textit{Fermi}-LAT light-curve showing at least a two-fold flux increase with respect to the average flux in the previous bins.
Analogously, the end of the `Flare' period was set to MJD 59190, since the flux drops significantly after that time.
No such high states of \btwo\ were recorded at $\gamma$ rays either between 2008 and the 2020 outburst or after that event.
In view of the discussion on the multi-band correlations (Section \ref{subsec:cross-correlations}), it is worth to highlight the following.
While most of the monthly bins during the $2010-2012$ period resulted in upper limits, 
as the time of the 2020 $\gamma$-ray flare got closer, the upper limits become less frequent and the average reconstructed flux slowly increases, up to the 2020 high state.
This indication of evolution on timescales of years can be traced in connection with the evolution of the emission in the other wavebands (Section \ref{subsec:Optical}).

The fast variability of the HE $\gamma$-ray flux during the 2020 flare was investigated by computing the weekly-binned light-curve during the `Flare' period (Figure \ref{fig:b2_LAT_weekly_optical}). The variability analyses performed on this light-curve, described in Section \ref{subsec:variability_weekly}, were used to identify a period of approximately 70 days surrounding the MAGIC observations, corresponding approximately to the double-peak structure at MJD $59110 - 59160$ in Figure \ref{fig:b2_LAT_weekly_optical}. In this period, the daily \textit{Fermi}-LAT light-curve was computed.
The time bins of the daily light-curve were determined by requiring that the bins in which the MAGIC observations fell were approximately centered on the MAGIC observing times, reported in Table \ref{tab:magic_table}.
Additionally, we constructed light-curves with adjustable-width time bins using the `adaptive-binning' method \citep{2012A&A...544A...6L}.
This method naturally adjusts the bin widths by requiring a constant relative flux uncertainty $\sigma_{F}/F$, so that low-activity periods are encapsulated within longer bins and high-activity periods are divided into shorter bins, without favoring any \textit{a priori} arbitrary timescale. Both the daily light-curve and the adaptive-binning with $\sigma_{F}/F \leq 0.3$ light-curve are shown in Figure \ref{fig:B2_flare_LC}.
With respect to the average flux above 100 MeV in quiescent state, i.e. $\left( 1.7\pm0.1 \right)\times 10^{-8}$ cm$^{-2}$ s$^{-1}$ (Table \ref{tab:summ_table_spectral_B2}), the daily flux on the day of the \textit{Fermi}-LAT high-state observation is $\left( 3.0\pm1.0 \right) \times 10^{-7}$ cm$^{-2}$ s$^{-1}$, hence indicating a daily flux increase by a factor $18 \pm 6$ in the HE $\gamma$-ray band.
The fast variability in the daily \textit{Fermi}-LAT light-curve can be exploited to constrain the size of the emission region responsible for the $\gamma$-ray flare, as discussed in Section \ref{subsec:variability}.

\subsection{\textit{Swift}}
\label{subsec:Swift_intro}
The \textit{Neil Gehrels Swift} satellite \citep{2004ApJ...611.1005G} is a rapidly slewing, multi-wavelength observatory primarily designed to observe gamma-ray bursts and their X-ray and optical/UV afterglows with its three instruments onboard: the Burst Alert Telescope \citep[BAT,][]{2005SSRv..120..143B} sensitive in the range $15-350\,$keV, the X-Ray Telescope \cite[XRT,][]{2005SSRv..120..165B}, sensitive to $0.2-10\,$keV photons, and the UV/Optical Telescope \cite[UVOT,][]{2005SSRv..120...95R} covering the $190-600\,$nm range.
Data from \textit{Swift}-BAT were not included in this work as the hard X-ray fluxes from blazars are generally below the BAT sensitivity for the typical exposure times of the order of several ks. Standard \textit{Swift} data-taking consists in simultaneous observations with its three telescopes.

Besides being part of a long-term monitoring by \textit{Swift}, \btwo\ was observed 8 times in 2020 as part of the MWL campaign, gathering total exposure times of $14.1\,\text{ks}$ for both \textit{Swift}-XRT and \textit{Swift}-UVOT. 
Additionally, the source lies in the field-of-view of the SAFE POINTING 3 observing mode.
The $\textit{Swift}$ safe pointings are predetermined locations on the sky that are observationally safe for UVOT and to which the spacecraft points when observing constraints do not allow observations of automated or pre-planned targets \citep{2004ApJ...611.1005G}.
When the spacecraft operates in safe pointing mode, only data from XRT and BAT are taken, whereas the exposure of UVOT is zero.
The first observations date back to 2005, whereas the latest ones we included were acquired in February 2024, allowing us to characterize in detail the long-term optical-to-X-ray SED evolution.

\subsubsection{\textit{Swift}-XRT}
\label{subsec:Swift_XRT_analysis}

The reduced XRT data products used for this study were downloaded from the UK \textit{Swift} Science Center\footnote{\url{https://www.swift.ac.uk/}} \citep{2009MNRAS.397.1177E}. Calibrated source and background files, along with the appropriate response files, were used as inputs to the spectral fitting package XSPEC v12.13.1 \citep{1996ASPC..101...17A} through PyXspec v2.1.2.

The spectra in the $0.3-10\,\text{keV}$ band were fitted with the {\fontfamily{cmtt}\selectfont tbabs} photon absorption model implemented in XSPEC folded with a PL (Eq. \ref{eq:PL}) and a BPL model of the form
\begin{equation}
\label{eq:BPL}
\frac{dN}{dE} = \left\{
              \begin{array}{ll}
                K \left( \frac{E}{E_{0}} \right)^{-\Gamma_{1, \text{BPL}}} & \quad E < E_{\text{break}} \\
                K E_{\text{break}}^{\Gamma_{2, \text{BPL}} - \Gamma_{1, \text{BPL}}} \left( \frac{E}{E_{0}} \right)^{-\Gamma_{2, \text{BPL}}} & \quad E \geq E_{\text{break}}
              \end{array}.
           \right.
\end{equation}
In the fit, the neutral hydrogen (HI) column density $N_\text{H}$ was fixed to the value reported in \cite{2013MNRAS.431..394W}, $N_\text{H} = 6.09\times10^{20}\,\text{atoms/cm}^{2}$, and we adopted the molecular abundancies reported in \cite{2000ApJ...542..914W}.
The Cash-statistic \citep{1979ApJ...228..939C} was employed in the fit procedures, the free parameters being the normalization and the spectral index (indices, in case of BPL model) of the source model.
The XSPEC implementation of the $F-$test\footnote{
\url{https://heasarc.gsfc.nasa.gov/xanadu/xspec/manual/node82.html}} was used to test whether the BPL model was preferred over the PL one on a statistical basis, the cut being fixed to $3\sigma$ significance level.

\begin{figure} [t] 
\centering
\begin{tabular}{c}
\includegraphics[width=0.99\columnwidth]{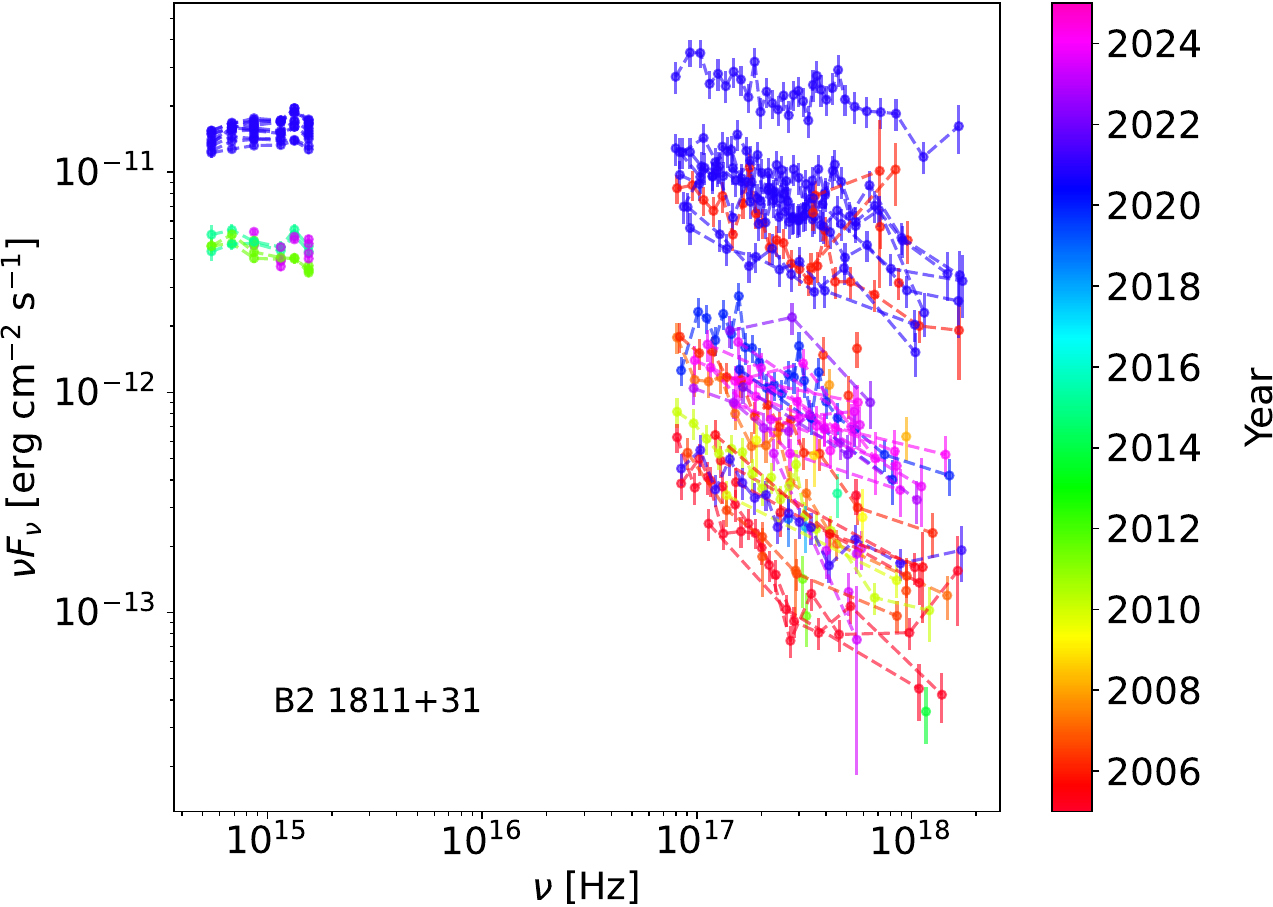}
\end{tabular}
\caption{Long-term evolution of the optical-to-X-ray SED of \btwo\ reconstructed from the \textit{Swift}-UVOT and \textit{Swift}-XRT observations.}
\label{fig:optical-to-X_SED_evolution_B2}
\end{figure}

The results of the spectral analyses of the \btwo\ \textit{Swift}-XRT observations taken during the 2020 $\gamma$-ray high-state period are reported in Table \ref{tab:results_XRT_B2}.
In 2 out of the 8 observations during the 2020 high state, the BPL model was preferred over the PL model for describing the X-ray spectrum of the source, with a spectral break at around $2-3\,$keV.
The X-ray spectral points were corrected for the ISM extinction effects, in order to reconstruct the intrinsic X-ray SED. The bin-by-bin correction factor was constructed by integrating the $e^{-\sigma_{\text{ISM}}(E) N_\text{H}}$ absorption term over each energy bin, using the HI column density value previously reported, where $\sigma_{\text{ISM}}(E)$ is the total photo-ionization cross-section of the ISM, normalized to the total hydrogen number density $N_\text{H}$ along the line-of-sight, in atoms cm$^{-2}$.

During the MAGIC observation period, two \btwo\ observations were carried out by \textit{Swift}-XRT, in MJD 59128 and MJD 59132. The spectra acquired in these observations yield X-ray fluxes of $\left( 11.0 \pm 0.6\right)$ and $\left( 11.4 \pm 0.6\right)$, in units of 10$^{-12}$ erg cm$^{-2}$ s$^{-1}$ and PL spectral indices $\Gamma_{\text{PL}}$ of $2.5 \pm 0.1$ and $2.6 \pm 0.1$, respectively. In Section \ref{sec:modeling}, we employed these measurements to gather insights into the particle population dominating the X-ray flux from the source.

The long-term evolution of the \btwo\ SED in the $0.3-10\,$keV energy range, corresponding to the $0.7 \times 10^{17} - 2.4 \times 10^{18}\,$Hz frequency range, is shown in Figure \ref{fig:optical-to-X_SED_evolution_B2}.
A variability in the X-ray flux by more than 2 orders of magnitude at $10^{18}\,\text{Hz}$ is observed.
Most of the highest-flux SEDs in Figure \ref{fig:optical-to-X_SED_evolution_B2} correspond to observations taken during the 2020 high state.
In addition, some high-state X-ray SEDs, shown in red in Figure \ref{fig:optical-to-X_SED_evolution_B2}, were computed from observations carried out during an X-ray flare in 2005. The available optical observations in 2005 from CRTS do not indicate an analogous high state in the optical R-band (Figure \ref{fig:B2_long_term_LC}).
Figure \ref{fig:optical-to-X_SED_evolution_B2} shows that during the 2020 flare the source exhibited a harder and brighter X-ray state compared to its low state. 
A similar trend is observed in HE $\gamma$ rays when comparing the average spectra between high and low states, as shown in Figure \ref{fig:LAT_preflare_flare_postflare} and discussed in Section \ref{subsec:Fermi_analysis}. The interpretation of this behavior, along with the classification of the high and low states of the source, is discussed in Section \ref{sec:source_classification}.

\subsubsection{\textit{Swift}-UVOT}
\label{subsec:Swift_UVOT_analysis}
The UVOT data were processed and analyzed using the {\fontfamily{cmtt}\selectfont uvotimsum} and {\fontfamily{cmtt}\selectfont uvotsource} tasks integrated in the HEASOFT 6.32 software package, along with the Swift/UVOTA Calibration Database (CALDB) files v20211108.
UVOT observations were performed using the optical \textit{v, b, u} and UV \textit{w1, m2, w2} photometric broad-band filters \citep{2008MNRAS.383..627P, 2010MNRAS.406.1687B}.
The {\fontfamily{cmtt}\selectfont uvotsource} tool performs aperture photometry on a single source and returns its count rates, magnitude and flux density accounting for instrumental effects embedded in the CALDB.
Source counts were extracted from a circular region of 5 arcseconds radius centered on the source, while background counts were derived from a circular region of 20 arcseconds radius in a nearby region free of contaminating sources.
To correct for Galactic interstellar extinction effects, the standard procedure described in \cite{1999PASP..111...63F} was employed, using $\text{E(B$-$V)} = 0.0431 \pm 0.0017$ \citep{2011ApJ...737..103S}.
The reconstructed magnitudes from the observations performed during the 2020 high state are reported in Table \ref{tab:summ_table_UVOT_obs_B2}.
The evolution of the long-term optical/UV SED is in Figure \ref{fig:optical-to-X_SED_evolution_B2} and it shows that during the 2020 high state the optical/UV flux experienced a two-fold increase with respect to the quiescent state.
In Section \ref{sec:source_classification}, we discuss the classification of the low and high states based on the selection of simultaneous UVOT and XRT data.

\subsection{Optical}
\label{subsec:Optical}
In this project, we include optical data from the following surveys, observatories and programs: Zwicky Transient Facility \cite[ZTF,][]{2019PASP..131a8002B}, Katzman Automatic Imaging Telescope \cite[KAIT,][]{2001ASPC..246..121F} at the Lick Observatory, Catalina Real-Time Transient Survey \cite[CRTS,][]{2009ApJ...696..870D}, Astronomical Observatory of the University of Siena\footnote{\url{https://www.dsfta.unisi.it/en/research/labs/astronomical-observatory}} ($0.3\,$m), W{\"u}rzburg Observatory\footnote{\url{https://schuelerlabor-wuerzburg.de/en/observatory/}} ($0.5\,$m) and the Tuorla blazar monitoring program\footnote{\url{https://users.utu.fi/kani/1m/}} \citep{nilsson2018}.
The Tuorla blazar monitoring program employed data from the 0.35\,m Kungliga Vetenskapsakademien (KVA) telescope until the end of 2019 and since 2022 it uses data from the Jean Orò Telescope (TJO)\footnote{\url{https://montsec.ieec.cat/en/joan-oro-telescope/}}.

The survey data have been extracted from the databases, while the dedicated observations of this source from Siena Observatory, W{\"u}rzburg Observatory and Tuorla blazar monitoring were analyzed using standard procedures of differential photometry \citep[e.g.][]{nilsson2018}. An aperture of $5\,$arcseconds was used for the photometry.   

In order to construct the long-term light-curve, we combined the survey data and dedicated observations, following the prescriptions of \cite{2024A&A...689A.119K}.
The dedicated observations were performed in the R-band, while the survey data were measured using many different filters.
In addition, the data from different telescopes can be affected by other instrumental systematic discrepancies (e.g. due to different apertures and different comparison stars).
In order to account for these effects, we used offsets of a few mJy, computed employing Tuorla blazar monitoring data as reference.
The shifts were determined using data from simultaneous or quasi-simultaneous nights.

The \btwo\ fluxes were corrected for host galaxy contribution, which was estimated to 0.015 mJy within aperture of 5 arcseconds, using the radial profile reported in \cite{2003A&A...400...95N}. Galactic extinction is 0.094 mag in R-band and the observed magnitudes were corrected accordingly \citep{2011ApJ...737..103S}.

The 2020 flare corresponds to a coherent high state in optical, X rays and $\gamma$ rays.
Moreover, the long-term optical light-curve shows that the 2020 high state occurred at the apex of the long-term average increasing trend in the optical band lasting for several years, since $2015-2016$.
After the high-state episode, the optical flux dropped with a decay time significantly shorter with respect to the timescale of the increasing trend leading to the 2020 high state.
The $\gamma$-ray long-term evolution indicates an analogous long-term rising-peaking-falling trend.

\subsection{Radio}
\label{subsec:Radio}
We included 15 GHz radio data from the AGN monitoring program of the $40\,\text{m}$ Telescope of the Owens Valley Radio Observatory (OVRO) and from the TELAMON program conducted with the Effelsberg $100\,\text{m}$ telescope at $14-17\,$GHz and $19-25\,$GHz.
The TELAMON flux densities are averages over multiple sub-frequencies within each band. For the $14-17\,$GHz band, the two sub-frequencies are $14.25$ and $16.75\,$GHz, whereas the four sub-frequencies within the $19-25\,$GHz band are $19.25$, $21.15$, $22.85$ and $24.75\,$GHz.
High-level data were provided by the respective instrument teams. The data analysis chains for the two radio telescopes are described in \cite{2011ApJS..194...29R} and \cite{2024A&A...684A..11E}, respectively for OVRO and TELAMON. 

\btwo\ was included in the OVRO monitoring program in 2009, thus providing a long-term radio coverage that allowed us to cross-correlate the long-term radio emission with those in the optical and $\gamma$-ray bands as reported in Section \ref{sec:variability_cross-correlations}.
Qualitatively, the long-term radio light-curve follows a different trend from the one displayed by the light-curves at higher frequencies. In particular, the two radio flares in 2017 and 2021, visible in the bottom panel of Figure \ref{fig:B2_long_term_LC}, have no trivial correspondence at higher frequencies.

\section{Variability and correlations}
\label{sec:variability_cross-correlations}

\subsection{Variability analysis}
\label{subsec:variability_intro}
The amplitude of the variations in time of the radiative emissions from AGNs can be orders of magnitude greater than in astrophysical sources such as stars and non-active galaxies.
The timescales of the fast coherent variations in the blazars emission in a given waveband are usually employed, through causality arguments, to constrain the size $R_\text{b}$ of the emission region dominating the radiative output in the same waveband. The observed timescales $t_{\text{var}}$ for variation in the radiation are longer than the light-crossing time, which results in 
\begin{equation}
    \label{eq:R_max}
    R_{\text{b}} \lesssim \frac{ c\, \delta_{\text{D}} \,t_{\text{var}}} {1+z}
\end{equation}
after accounting for Doppler boosting between the observer frame and the reference frame comoving with the emission region and for cosmological effects through the source redshift $z$.
The relativistic Doppler factor is defined as
\begin{equation}
    \delta_{\text{D}} = \left[ \Gamma (1 - \beta \cos{\theta}) \right]^{-1}
    \label{eq:Doppler_factor}
\end{equation}
and governs the transformation of photon energies between the frames with relative Lorentz factor $\Gamma= 
\left( 1 - \beta^{2} \right)^{-1/2}$, where $\vec{\beta}c$ denotes the velocity of the moving emission region and $\theta$ is the angle formed by $\vec{\beta}c$ with the line of sight from the observer.

\subsubsection{Short-timescale variability}
\label{subsec:variability}
In the literature, constraints on the emission region size have been estimated from the fast variability in many different wavebands, such as X rays, HE $\gamma$ rays and VHE $\gamma$ rays \citep[e.g.][ respectively]{2024A&A...685A.117M, 2013A&A...555A.138F, 2007ApJ...664L..71A}.
In this project, we use the method described in \cite{2011A&A...530A..77F} to infer the short-timescale variability of the HE $\gamma$-ray flux during the $\gamma$-ray high-state period and to quantify the significance of the estimate.
The method consists in scanning the \textit{Fermi}-LAT daily light-curves in Figure \ref{fig:B2_flare_LC} to find the minimum doubling/halving time $\tau$, defined from
$    F(t) = F(t_{0}) \, 2^{-(t - t_{0})/ \tau} $,
where $t$ and $t_{0}$ are the centers of two consecutive time bins in which the source fluxes $F(t)$ and $F(t_{0})$ were reconstructed without ending up into upper limits. The significance of the reconstructed $\tau$ was computed as $S = |F(t) - F(t_{0})| / \sqrt{ \sigma_{F(t)}^{2} + \sigma_{F(t_{0})}^{2} }$, where $\sigma_{F(t)}$ and $\sigma_{F(t_{0})}$ are the uncertainties on $F(t)$ and $F(t_{0})$, respectively.

The resulting variability timescale with the highest significance levels are reported in Table \ref{tab:variability_table_B2}.
The MWL follow-up campaign on \btwo\ was organized after the $\textit{Fermi}$-LAT observation of elevated $\gamma$-ray emission from the source.
Indeed, the days in which the daily flux yielded the most significant variability timescales closely correspond to those surrounding the $\textit{Fermi}$-LAT detection of high state.
The resulting variability timescale with the highest significances are $t_{\text{var}} \approx \left(3 - 6\right) \,\text{h}$, so the size of the emission region dominating the $\gamma$-ray flux with relativistic Doppler factor $\delta_\text{D}$ has to be smaller than $\text{R}_{\text{max}} \approx \left( 3 - 6 \right) \times 10^{14}\, \delta_\text{D}\, \text{cm}$.
Variability timescales of the order of few hours are compatible with those found for other TeV blazars in flaring state \citep[e.g.][]{2021MNRAS.507.1528A, 2024A&A...684A.127A}.

\begin{table}
\small
\caption{Results of the short-timescale variability analysis of the HE $\gamma$-ray emission of \btwo.}
\label{tab:variability_table_B2}\begin{tabular}{ccccccc}
\hline
\multicolumn{1}{c}{$t_{0}$} &
\multicolumn{1}{c}{$t$}&
\multicolumn{1}{c}{$F (t_{0})\,\times\,$10$^{-8}$} &
\multicolumn{1}{c}{$F (t)\,\times\,$10$^{-8}$} &
\multicolumn{1}{c}{$\tau$} &
\multicolumn{1}{c}{$S$} \\ 
\multicolumn{1}{c}{[MJD]} &
\multicolumn{1}{c}{[MJD]} &
\multicolumn{1}{c}{[ph cm$^{-2}$s$^{-1}$]} &
\multicolumn{1}{c}{[ph cm$^{-2}$s$^{-1}$]} &
\multicolumn{1}{c}{[h]} &
\multicolumn{1}{c}{} \\
\hline
59122.9 & 59123.9 & $7.8\pm4.4$ & $30\pm10$ & 6.1 & $2.1$ \\
59123.9 & 59124.9 & $30\pm10$ & $2.1\pm1.4$ & 3.1 & $2.8$ \\
\hline
\end{tabular}
\tablefoot{Only the two variability timescales with the highest significance $S$ are reported. If $F (t) > F (t_{0})$, the corresponding $\tau$ is a doubling time, otherwise $\tau$ corresponds to a halving time.}
\end{table}

\subsubsection{Variability in timescales of several days}
\label{subsec:variability_weekly}

In this section, we present the analysis of the HE $\gamma$-ray flux variability in timescales of several days in a period of approximately 250 days surrounding the \textit{Fermi}-LAT high-state observation.
Indeed, as shown in the long-term MWL light-curve in Figure \ref{fig:B2_long_term_LC}, \btwo\ persisted in a relatively high state (with respect to the average flux several years prior) from April 2020 ($\text{MJD}\approx58940$) to December 2020 ($\text{MJD}\approx59190$), thus from several months before the \textit{Fermi}-LAT high-state observation. 
This period was denoted as `Flare' in Table \ref{summ_table_b2}.

Figure \ref{fig:b2_LAT_weekly_optical} shows the $\textit{Fermi}$-LAT light-curve with weekly bins and the optical R-band light-curve in the `Flare' period.
The \textit{Fermi}-LAT light-curve with bins of 7 days was computed using the method described in Section \ref{subsec:Fermi_analysis}.
Visual inspection of the two light-curves qualitatively indicates a common trend between the flux evolution in the two energy bands.
The cross-correlation between the optical and HE $\gamma$-ray fluxes is quantitatively evaluated in Section \ref{subsec:cross-correlations}.
The light-curves show repeating rising and falling trends.
We employed a double-exponential function 
\begin{equation}
    \label{eq:double_exponential}
    F(t) = A \cdot 
    \begin{cases}
        2^{ \left( \frac{t - t_\textrm{peak}}{ t_\textrm{rise}} \right) } \qquad \qquad \textrm{if} \qquad t \leq t_\textrm{peak}\\
        2^{ \left( \frac{t_\textrm{peak}-t}{ t_\textrm{decay}} \right) } \qquad \qquad \textrm{if} \qquad t > t_\textrm{peak}
    \end{cases}
\end{equation}
to fit the peaks of the \textit{Fermi}-LAT weekly-binned light-curve around the three highest-flux bins, for which the optical R-band light-curve shows similar trends.
In the time interval $\text{MJD}\,58940\,-\,58970$, the fit yielded $t_\textrm{rise}=\left( 5.6 \pm 1.6 \right)$ d and $t_\textrm{decay}=\left( 4.5 \pm 1.6 \right)$ d for the bump peaking at $t_\textrm{peak}=\left( 58962 \pm 2 \right)$ MJD.
Analogously, the fit of the light-curve in the $\text{MJD}\,59020\,-\,59050$ time interval yielded $t_\textrm{rise}=\left( 7.4 \pm 2.1 \right)$ d and $t_\textrm{decay}=\left( 6.2 \pm 2.0 \right)$ d, with $t_\textrm{peak}=\left( 59042 \pm 2 \right)$ MJD.
The double-peak structure in the $\text{MJD}\,59110-59160$ time interval was fitted with the sum of two double-exponential functions (Eq. \ref{eq:double_exponential}).
We retrieved estimates of $t_\textrm{rise} = \left( 18 \pm 4 \right)$ d for the first peak and $t_\textrm{decay} = \left( 11 \pm 3 \right)$ d for the second one, while the other temporal parameters of the two double-exponential function were poorly constrained.
Figure \ref{fig:B2_flare_LC} shows the MWL light-curve in the $\text{MJD}\,59110-59160$ period.
In Section \ref{sec:modeling} we discuss how these estimates can act as guide for modeling the broad-band SED during the 2020 high state.

\subsection{Intra-band correlations}
\label{subsec:intra-band_correlations}

In this section, we discuss the intra-band correlations between the flux $F$ and photon index $\Gamma$ at X rays and at HE $\gamma$ rays.

\subsubsection{X rays (\textit{Swift}-XRT)}
\label{subsec:intraband_correlations_X_rays}

\begin{figure}
\centering
\begin{tabular}{c}
\includegraphics[width=0.92\columnwidth]{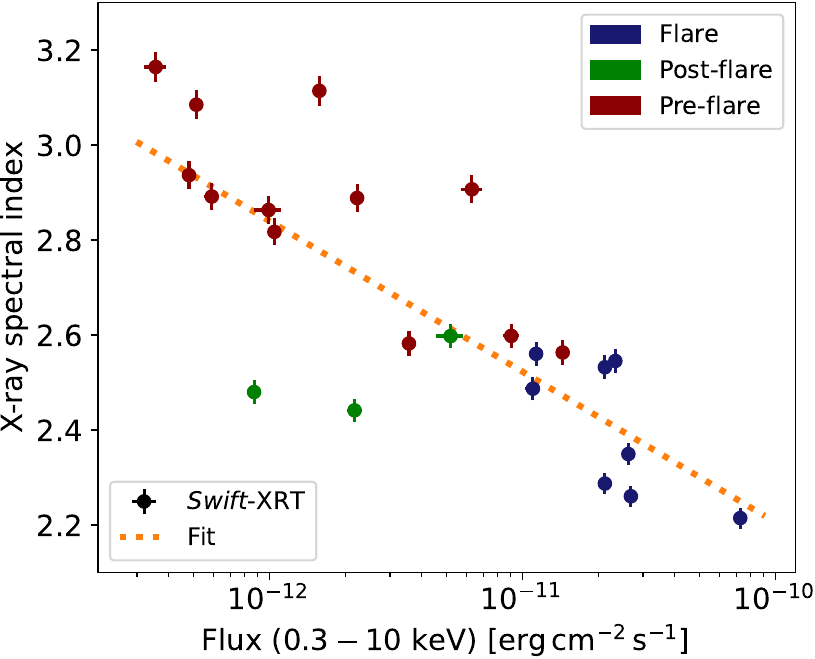}
\end{tabular}
\caption{Scatter plot of the intrinsic power-law spectral index $\Gamma_{\text{X}}$ and flux $F_{\text{X}}$ in the $0.3-10\,\text{keV}$ energy range from the long-term \textit{Swift}-XRT observations of \btwo. The observations carried out the `Pre-flare', `Flare' and `Post-flare' periods are marked in red, blue and green, respectively.
The orange line marks the fit of the full dataset with Eq. \ref{eq:X}.}
\label{fig:harder-when-brighter}
\end{figure}

Figure \ref{fig:harder-when-brighter} shows the scatter plot of the photon index $\Gamma_{\text{X}}$ and flux $F_{\text{X}}$ in the $0.3-10\,$keV energy range from the \textit{Swift}-XRT data presented in Section \ref{subsec:Swift_XRT_analysis}.
We primarily tested the correlation between $\Gamma_{X}$ and $F_{X}$ separately during the `Pre-flare', `Flare' and `Post-flare' periods using the Pearson coefficient $r$, a statistical measure of linear correlation.
The statistical uncertainties were included in the computation of both the correlation coefficient and significance through simulations, as described in Appendix \ref{sec:appendix_intraband}.
We find that during the `Pre-flare' (`Flare') period the Pearson coefficient between $\Gamma_{\text{X}}$ and $\log_{10} F_{\text{X}}$ resulted to be $r=-0.75$ ($r=-0.71$), with $p\text{-value}=4\times10^{-3}$ ($p\text{-value}=2\times10^{-2}$), corresponding to a significance of $2.9\sigma$ ($2.3\sigma$) for the null hypothesis of non-correlation.
Therefore, $\log_{10} F_{\text{X}}$ is linearly anti-correlated with $\Gamma_{\text{X}}$ with a confidence level of 95\% in both the `Pre-flare' and `Flare' periods.
However, the limited number of \textit{Swift}-XRT observations in the `Post-flare’ period prevents a reliable statistical analysis of the source behavior during this phase.
Over the full dataset, which includes data from all periods combined, we find $r = -0.77$ with $p\text{-value} = 4 \times 10^{-6}$ (significance of $4.6\sigma$).
The trend was fitted with a linear function between $\Gamma$ and $\log_{10}F$\begin{equation}
    \Gamma = p_{0} + p_{1} \log_{10} \left(F/F_0\right)\,,
    \label{eq:X}
\end{equation} with $F_{0}=1\,\text{erg}\,\text{cm}^{-2}\,\text{s}^{-1}$, yielding best-fit values $p_{0} = -1.1 \pm 0.6$ and $p_{1} = -0.33 \pm 0.05$.

The analysis reported here indicates that the long-term evolution of the X-ray spectrum of the source follows a harder-when-brighter trend, i.e. higher X-ray fluxes correspond to harder spectra. This trend has been observed in the evolution of the X-ray emission of several IBLs and HBLs \citep[e.g. Mrk\,421,][]{2015A&A...576A.126A} and it can be explained in terms of the dynamics of the electron population that is dominating the emission in the X-ray band. 
Indeed, for HBLs and IBLs with $\nu_{\text{s}}$ close to $10^{15}$ Hz (Section \ref{sec:source_classification}), the $0.3-10\,$keV energy band typically hosts the falling edge of the synchrotron SED bump. Therefore, this radiation is emitted by the highest-energy electrons in the jet, above the break of the particle spectrum.

Acceleration mechanisms result in the injection of particles in the emission regions with a hard spectrum, e.g. $\frac{dN}{dE} \propto E^{-2}$ from first-order Fermi acceleration.
As a consequence, after an increase in the particle injection rate, the X-ray spectrum hardens and the X-ray flux increases. In other words, high X-ray fluxes correspond to hard X-ray spectra.

The highest-energy electrons primarily cool through synchrotron radiation emission. Moreover, synchrotron cooling proceeds faster with increasing particle energy (Section \ref{sec:modeling}). 
Therefore, after the injection of accelerated particles with a hard spectrum, synchrotron cooling simultaneously softens the high-energy tail of the particle spectrum and lowers the density of particles with sufficient energy to radiate photons in the $0.3-10\,\text{keV}$ band \citep[e.g.][]{2022A&A...658A.173T}. This implies that low X-ray fluxes correspond to soft X-ray spectra.

\subsubsection{High-energy $\gamma$ rays (\textit{Fermi}-LAT)}
\label{subsec:intraband_correlations_HEgamma_rays}

\begin{figure} [t] 
\centering
\begin{tabular}{c}
\includegraphics[width=0.85\columnwidth]{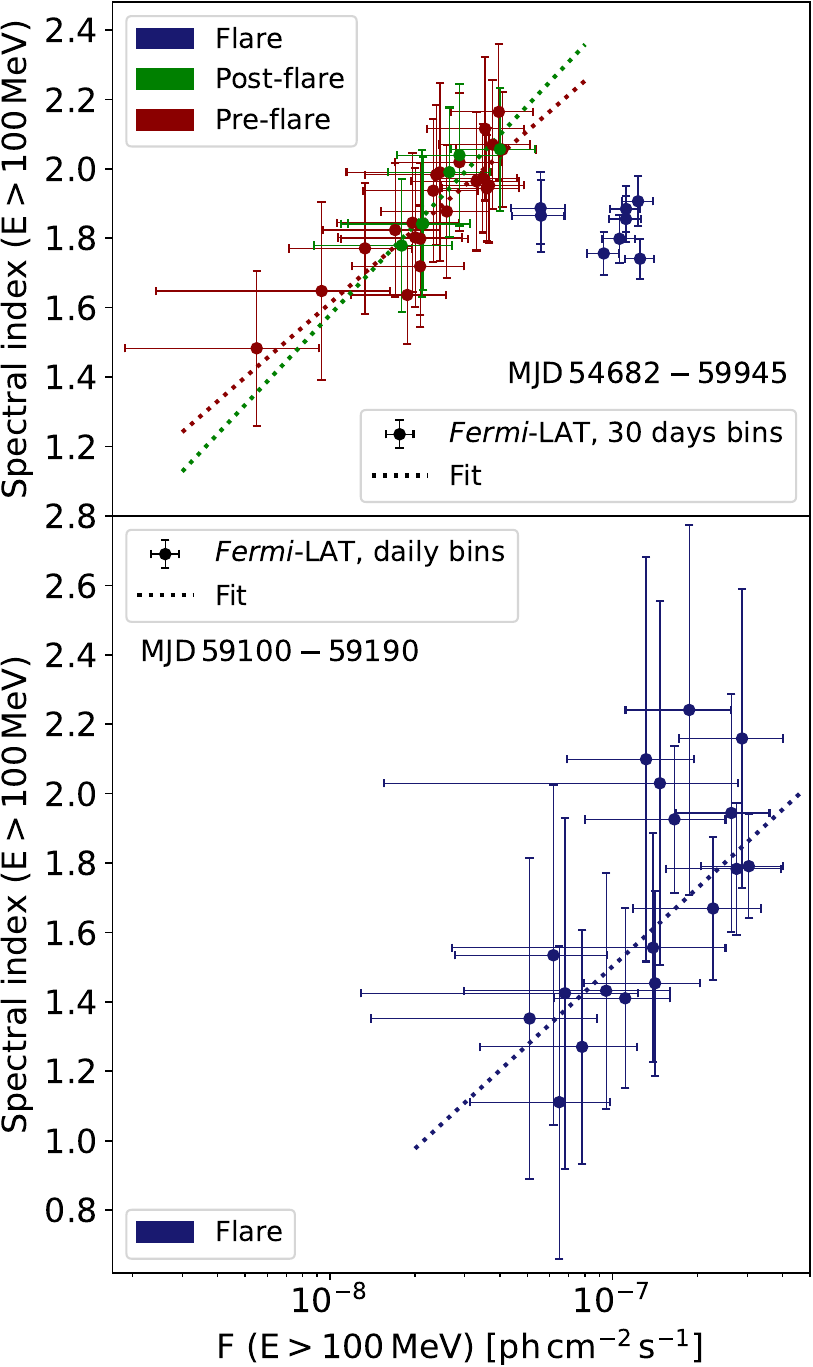}
\end{tabular}
\caption{
Scatter plot of the PL spectral index $\Gamma_{\gamma}$ and flux $F_{\gamma}$ in the HE $\gamma$-ray data. In the top (bottom) panel, $\Gamma_{\gamma}$ and $F_{\gamma}$ derive from the \textit{Fermi}-LAT light-curve with 30-day bins (daily bins) shown in Figure \ref{fig:B2_long_term_LC} (Figure \ref{fig:B2_flare_LC}).
The dotted lines mark the fits using Eq. \ref{eq:X}.}
\label{fig:Fermi_LAT_correlations}
\end{figure}

In Section \ref{subsec:Fermi_analysis}, we pointed out that during the 2020 high state the source exhibited on average a harder and brighter HE $\gamma$-ray state compared to the low states, as shown in Figure \ref{fig:LAT_preflare_flare_postflare}.
Conversely, in this section we provide evidence that the source spectral properties at HE $\gamma$ rays on daily timescales during the 2020 high state and on monthly timescales during the low states show hints of softer-when-brighter trends instead.

Figure \ref{fig:Fermi_LAT_correlations} shows the scatter plot of the photon index $\Gamma_{\gamma}$ and flux $F_{\gamma}$ in the $100\,\text{MeV}-1\,\text{TeV}$ energy range from the \textit{Fermi}-LAT light-curves with bins of 30 days in the top panel and with daily bins in the bottom panel.
Only the temporal bins in which the $TS$ of the source was found to be above 25 were included. 
The top panel shows that the 2020 high state is well separated in the $\Gamma_{\gamma}-F_{\gamma}$ plane from the low states, i.e. `Pre-flare' and `Post-flare' periods. Nonetheless, in two temporal bins within the `Flare' period with flux $F_{\gamma}\approx5.5\times10^{-8}\,\text{ph}\,\text{cm}^{-2}\,\text{s}^{-1}$, the source likely transitioned between the high and low states.
By employing the Pearson coefficient $r$, we analyzed the linear correlation between $\Gamma_{\gamma}$ and $\log_{10} F_{\gamma}$ during the low states using the light-curve with 30-day bins from the `Pre-flare' and `Post-flare' periods (top panel). 
Conversely, the correlation during the 2020 high state (bottom panel) was investigated using the daily light-curve in Figure \ref{fig:B2_flare_LC}.
The statistical uncertainties were included in the analysis as described in Appendix \ref{sec:appendix_intraband}. 
In addition, in each state the trends were fitted using Eq. \ref{eq:X}, with $F_{0}=1\,\text{ph}\,\text{cm}^{-2}\,\text{s}^{-1}$.
The results are reported in Table \ref{tab:summ_table_Fermi_correlations}.

A positive linear correlation with $r=0.41$ is found with $p\text{-value} = 0.033$ (significance of $2.1\sigma$) during the `Pre-flare', thus indicating with a $\sim$95\% confidence level a softer-when-brighter trend, i.e. higher fluxes correspond to softer spectra.
In the `Flare' and `Post-flare' periods, respectively, we find positive $r$ values equal to $r=0.32$ and $r=0.25$, with $p$-values of $0.10$ and $0.32$ ($1.7\sigma$ and $1.0\sigma$). The lower significance values compared to the `Pre-flare' period are likely due to the lower number of bins and to the higher influence of statistical uncertainties.

Although the correlation is statistically significant above the $2\sigma$ level only during the Pre-flare period, it is noteworthy that in all three periods we find hints of positive correlation between the flux $F_\gamma$ and the photon index $\Gamma_\gamma$. This suggests that the softer-when-brighter trend could be an intrinsic feature of the source spectral evolution at HE $\gamma$ rays within a given activity state, i.e. high or low. However, due to the limited statistics, a definitive conclusion cannot be drawn from the current dataset.

In the following, we report two key aspects regarding the energy range and main cooling mechanism of the electrons responsible for the source HE $\gamma$-ray emission. We then discuss the interpretation of the softer-when-brighter trend at HE $\gamma$ rays, in the context of the processes occurring within the jet.

The energy range of the electron distribution dominating the emission in the HE $\gamma$-ray band is different compared to the one responsible for the X-ray emission (Section \ref{subsec:intraband_correlations_X_rays}).
Figure \ref{fig:LAT_preflare_flare_postflare} shows that the peak energy of the SED high-energy component is likely below $1\,\text{GeV}$ during the low states and it shifts at energies above tens of GeV during the high state.
In both activity states, the emission at energies between $100\,\text{MeV}$ and few GeV is mainly produced by particles with energies around or below the spectral break of the particle distribution \citep{2010ApJ...710.1271A}.

Moreover, for electrons in this energy range, the timescales for the escape from the emission region are shorter than those for cooling through synchrotron radiation emission (Section \ref{sec:modeling}). 
This implies that synchrotron emission is not the dominant cooling mechanism for electrons in this energy range. 
Therefore, the spectral properties of the source emission at HE $\gamma$ rays are primarily determined by the processes of acceleration and escape, rather than from synchrotron cooling \citep{2022A&A...658A.173T}.

Although this is not the first indication of softer-when-brighter trends in the spectral evolution at HE $\gamma$ rays of a blazar \citep[e.g.][]{2022MNRAS.515.2633P, 2023MNRAS.524.4333D}, the interpretation of this effect is still unclear.
Theoretical studies \citep[e.g.][]{2019ApJ...887..133B} indicate that acceleration processes tend to lead to harder-when-brighter trends in HE $\gamma$ rays, rather than softer-when-brighter ones.
One possibility is that softer-when-brighter trends may result when the total HE $\gamma$-ray flux of the source is the sum of the radiation from different emission components, e.g. SSC and external Compton \citep{2024ApJ...974..233K}, or from multiple emission regions.
In the case of \btwo, in Section \ref{sec:modeling} we investigate the influence of multiple emission regions to the broad-band emission of the source during the 2020 high state.
However, a quantitative investigation of the softer-when-brighter trends through a time-dependent model accounting for multiple emission regions \citep[e.g.][]{2019ApJ...887..133B, 2024ApJ...974..233K} is beyond the scope of this work.

\begin{figure*}
\centering
 \subfloat[zDCF of the long-term HE $\gamma$-ray and R-band light-curves.]
 {\includegraphics[width=0.95\columnwidth]{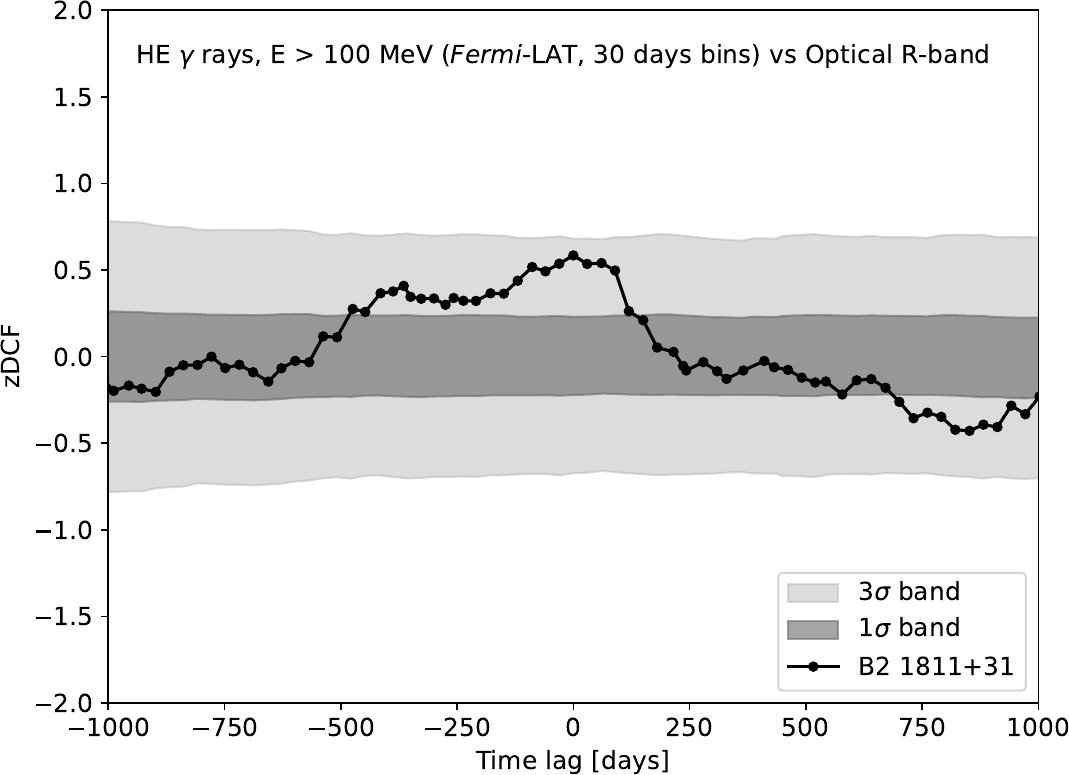} \label{fig:cross_correlation_B2_gamma_optical}}
 \hfill
 \subfloat[Correlation with 95\% CL of $\gamma$-ray and R-band light-curves.]{\includegraphics[width=0.94\columnwidth]{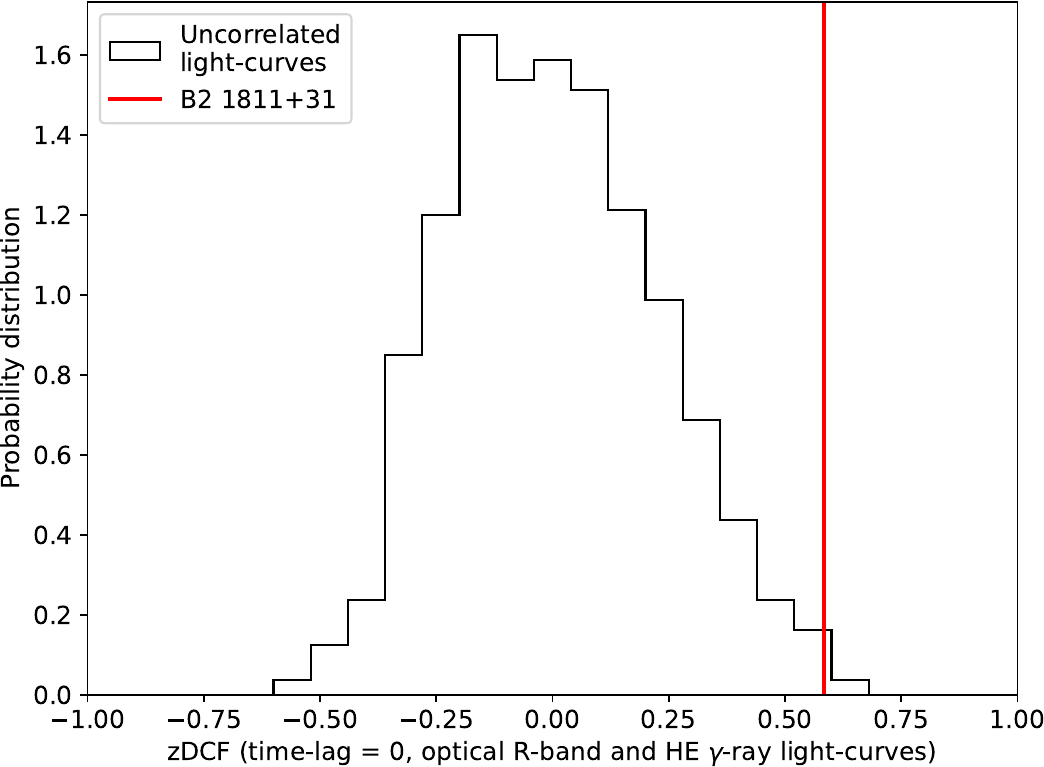} \label{fig:cross_correlation_B2_significance_gamma_optical}}

 \subfloat[zDCF of the long-term HE $\gamma$-ray and 15 GHz light-curves.]{\includegraphics[width=0.95\columnwidth]{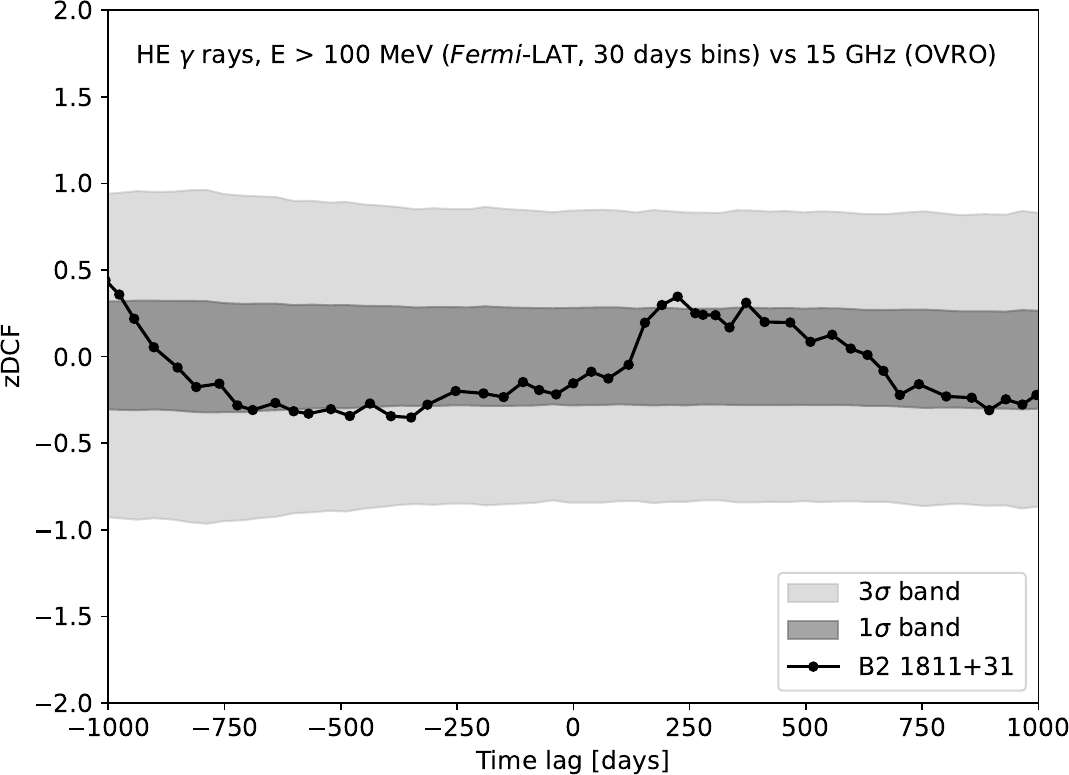} \label{fig:cross_correlation_B2_gamma_radio} }
 \hfill
 \subfloat[zDCF of the long-term R-band and 15 GHz light-curves.]{\includegraphics[width=0.95\columnwidth]{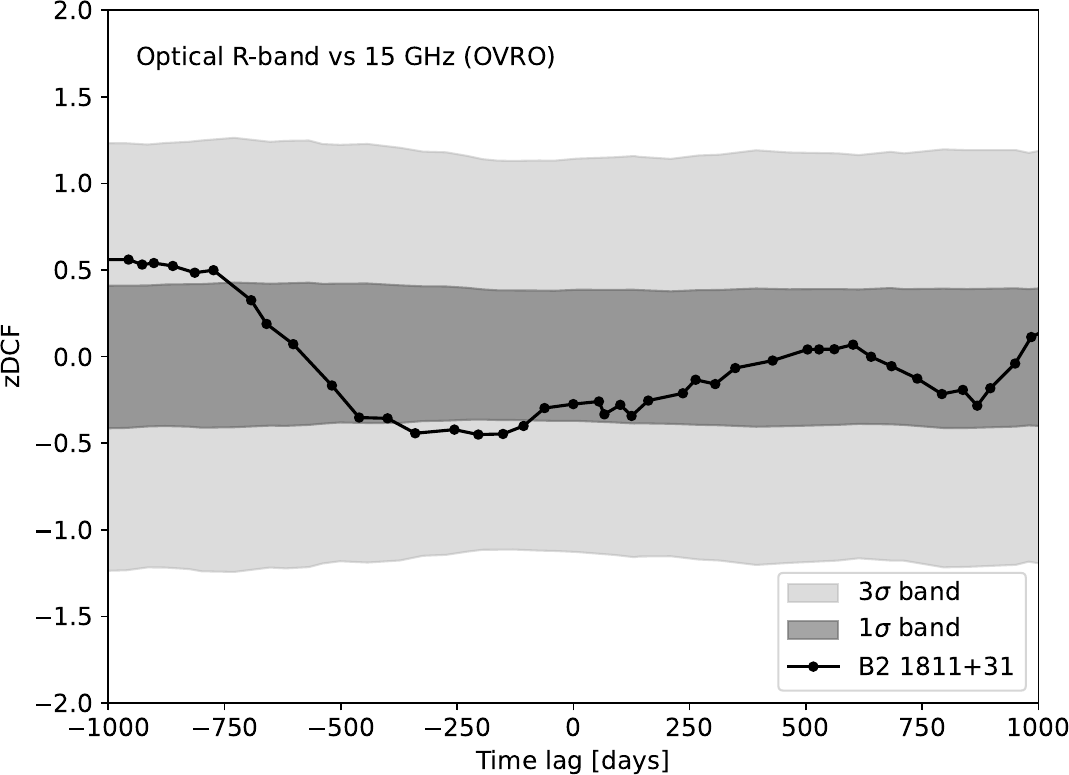}
 \label{fig:cross_correlation_B2_optical_radio} } 
 \\

 \caption{Summary of the zDCF analyses on the long-term \btwo\ light-curves: (a) HE $\gamma$-ray and R-band, (c) HE $\gamma$-ray and radio, (d) optical R-band and radio.
 The $1\sigma$ and $3\sigma$ bands are indicated in dark and light gray, respectively.
 Panel (b) shows the distribution of the reconstructed zDCF at zero time-lag for $10^{3}$ simulated uncorrelated HE $\gamma$-ray and optical light-curves, whereas the red line marks the value reconstructed with the corresponding observed \btwo\ light-curves in Figure \ref{fig:B2_long_term_LC}.}
 \label{fig:B2_cross_correlation}
\end{figure*}

\subsection{Multi-band correlations}
\label{subsec:cross-correlations}

Correlations between the observed fluxes at different wavelengths have been long employed to infer insights on the dynamical processes acting in AGN, for instance yielding evidence for BLR stratification \citep{1991ApJ...366...64C, 1991ApJ...368..119P}.
In recent years, multi-band cross-correlations of blazar emissions from radio to VHE $\gamma$ rays were investigated to distinguish the signatures of possible multiple emission components in the jet. These indications are then interpreted as guide for the spectral energy distribution modeling \citep{2016A&A...593A..98L}.

For this task, we employed the z-transformed Discrete Correlation Function \citep[zDCF,][]{1997ASSL..218..163A}, which is optimized for sparse and unevenly sampled astronomical time-series. The zDCF provides a more robust method for evaluating the delay-dependent multi-band cross-correlations than the Discrete Correlation Function \citep[DCF, ][]{1988ApJ...333..646E}, due to the utilization of equal-population binning and Fisher's \textit{z}-transform.

The significance of the zDCF is estimated by computing the zDCF of populations of simulated uncorrelated light-curves sharing the same power spectral density (PSD) as the observed ones.
Light-curves were generated with a Python 3 implementation\footnote{\url{https://github.com/cerasole/DELightcurveSimulation}} of the DELCgen package \citep{2016ascl.soft02012C}, which allows for the generation of light-curves using both \cite{1995A&A...300..707T} (T\&K) and \cite{2013MNRAS.433..907E} (EM) methods.
We decided to use the latter, as in the EM method the fluxes of the simulated light-curves could be generated from the probability density function of the fluxes of the observed light-curve. In the original T\&K method, no constraints on the simulated flux distribution are imposed, thus allowing for the generation of light-curves with bins of negative fluxes. For this reason, in order to produce light-curves that are both more physical and more resembling the observed ones, we employed the EM method.

We computed the cross-correlations of the long-term HE $\gamma$-ray, optical R-band and 15 GHz radio light-curves.
The PSDs of these long-term light-curves were fitted with power-law functions in the form $\text{PSD}(\omega) = A \omega^{-\beta}$.
The best-fit index for the $\gamma$-ray light-curve is $\beta_{\gamma} = 1.1$, in agreement with \cite{2020ApJS..250....1T} and references therein, which report typical values between 1.1 and 1.6 for the PSD indices for the \textit{Fermi}-LAT light-curves of a sample of 104 blazars.
The optical R-band light-curve yields a best-fit power index of $\beta_{\text{R}} = 1.5$, whereas the 15 GHz radio light-curve yielded $\beta_{15\,\text{GHz}} = 2.3$. These PSD power-law indices for the optical and radio AGN light-curves are compatible with those reported in \cite{2016A&A...593A..98L} for a sample of 32 VHE $\gamma$-ray emitting BL Lac objects.
For each long-term cross-correlation to be tested, $10^{3}$ pairs of uncorrelated light-curves were generated with PSD modeled as power-law with the corresponding indices and with the same flux distributions as the observed ones.
The results of the zDCF analysis are shown in Figure \ref{fig:B2_cross_correlation}.

Positive correlation with no delay between the long-term $\gamma$-ray and R-band light-curves with a confidence level of 95\% is established, as Figures \ref{fig:cross_correlation_B2_gamma_optical} and \ref{fig:cross_correlation_B2_significance_gamma_optical} show.
This trend is generally seen for LBLs, IBLs and FSRQs \citep{2016A&A...593A..98L} and it is related to the peak frequency ranges of the two main bumps characterizing the broad-band SED.
This evidence can be interpreted in terms of the same emission regions contributing significantly to the fluxes in the two energy bands.
On the other hand, the radio emission shows neither correlated flares nor correlated long-term trends with the emissions in the optical band and at HE $\gamma$ rays, as shown in Figures \ref{fig:cross_correlation_B2_gamma_radio} and \ref{fig:cross_correlation_B2_optical_radio}.
The absence of correlated trends between the long-term radio and optical R-band light-curves is not usual among BL Lacs, as shown in \cite{2016A&A...593A..98L}.
This evidence suggests that the bulk of the radio emission may originate from different zones from those dominating the radiative output in the optical band and at HE $\gamma$ rays.
The zones responsible for the most of the radio emission are expected to be larger, less relativistic and located further along the jet than those responsible for the bulk of the emission at higher frequencies.

\section{Source classification}
\label{sec:source_classification}

BL Lacs can be empirically classified according to the frequency of the SED low-energy bump into LBLs, IBLs and HBLs, as described in Section \ref{sec:intro}. 
During high-state periods, the synchrotron peak frequency can shift significantly \citep{1998ApJ...492L..17P} and the high-state classification can be different from the low-state one \citep{1999ApJ...515L..21B}.
In this section, we deal with the classification of the source during both the low and high radiative states.

\begin{figure}
\centering
\includegraphics[width = 0.95\columnwidth]{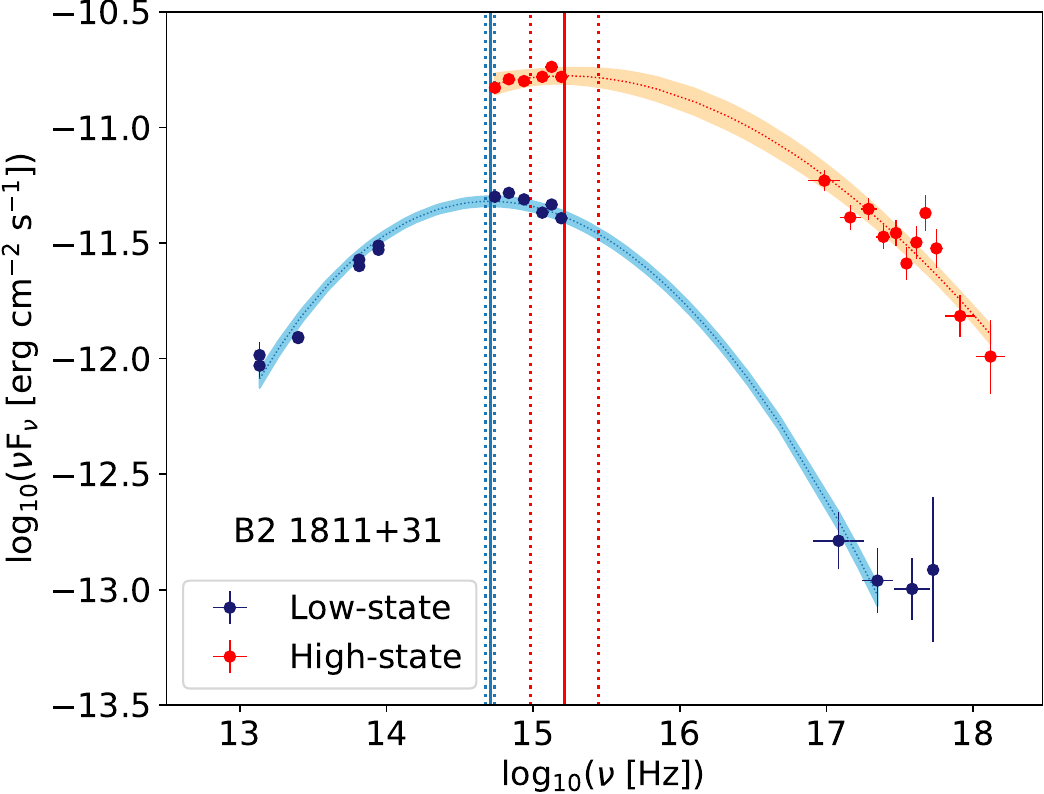}
\caption{
Fit of the infrared-to-X-ray SEDs characteristic of the \btwo\ low and high states with Eq. \ref{eq:logparabola}, in blue and red respectively.
The spectral points in the optical/UV range were reconstructed from \textit{Swift}-UVOT observations, whereas the X-ray ones derive from \textit{Swift}-XRT data. 
Infrared points are WISE data.
The shaded areas correspond to the $1\sigma$ band fit uncertainties, while the best-fit synchrotron peak frequencies and their uncertainties are marked by the solid and dotted vertical lines.
}
\label{fig:B2_logparabola}
\end{figure}

\btwo\ is classified as IBL in \cite{1999ApJ...525..127L}.
Figure \ref{fig:optical-to-X_SED_evolution_B2} shows the evolution of the optical-to-X-ray SED from 2005 to 2024 using \textit{Swift}-UVOT and \textit{Swift}-XRT data. 
As the variability of the optical-to-X-ray SED is remarkable, we focused on simultaneous observations in order to select source states to be classified.
The \textit{Swift}-UVOT and \textit{Swift}-XRT observations chosen as characteristic for the \btwo\ low state are those carried out on May $13-24$, 2011 ($\text{MJD}\,55694-55705$).
In order to increase the statistics and have more precise flux estimation, we included also the \textit{Swift}-UVOT and \textit{Swift}-XRT observations taken in January 2015, as the spectral properties in these observations were fully compatible with the 2011 observations. For the low state, in order to have coverage also in infrared, we included temporally close data taken from the Wide-field Infrared Survey Explorer \citep[WISE,][]{2010AJ....140.1868W} at 3.4, 4.6, 12 and 22 $\mu$m wavelengths (W1, W2, W3, W4 filters). The considered observations were performed in March 2010 and the high-level data were downloaded from the SED Builder tool\footnote{\url{https://tools.ssdc.asi.it/SED/}} provided by the ASI Space Science Data Center (SSDC). In addition, they connect smoothly to the optical/UV spectral points derived from the \textit{Swift}-UVOT observations characterizing the source low state, so we decided to include them as characteristic of the low state.
For the high state, we selected data from simultaneous \textit{Swift}-XRT and \textit{Swift}-UVOT data from the 2020 high state (Tables \ref{tab:results_XRT_B2} and \ref{tab:summ_table_UVOT_obs_B2}, respectively).
The synchrotron peak frequency was estimated by fitting the infrared-to-X-ray SED with a quadratic function in the $\log_{10} \nu - \log_{10} \nu F_{\nu}$ plane\begin{equation}
    \nu F\left(\nu\right) = f_{0} \, 10^{-b\, (\log_{10} \left( \nu/\nu_{\text{s}} \right))^{2} }\,,
    \label{eq:logparabola}
\end{equation}
where $\nu_{s}$ is the synchrotron peak frequency, $f_{0}$ is a normalization constant and $b$ is the coefficient of the second-order term in $\log_{10} \nu$.
The resulting $\nu_{\text{s}}$ values for the low and high states are
\begin{enumerate}
    \item Low state: $\log_{10} \left( \nu_{\text{s}} / \text{Hz} \right)= 14.71 \pm 0.03$
    \item High state: $\log_{10} \left( \nu_{\text{s}} / \text{Hz} \right)= 
    15.21 \pm 0.23$
\end{enumerate}
The fits of the selected data are shown in Figure \ref{fig:B2_logparabola}. The quiescent behavior is compatible with the one reported in \cite{1999ApJ...525..127L} for this source. Instead, the high-state classification is borderline between IBL and HBL, as the reconstructed $\nu_{\text{s}}$ is slightly above $10^{15}$ Hz.
Thus, a significant shift of the synchrotron peak frequency during the 2020 high state with respect to the quiescent state is established.

The TeV blazar 1ES\,1215+303, discovered at VHE by MAGIC \citep{2012A&A...544A.142A}, has shown several MWL similarities with \btwo, such as the long-term optical-gamma-ray flux increase, presented in Section \ref{subsec:Optical}, and the shift in the synchrotron peak frequency during high states \citep{2020ApJ...891..170V}. 
As \btwo, during low states 1ES\,1215+303 is classified as an IBL, whereas it shows HBL behavior during high-state episodes.
These similarities indicate common physical conditions in the jets of the two blazars. Moreover, they provide evidence for sources crossing the boundaries of the standard partition of BL Lacs based on the synchrotron peak frequency.

In the following, we provide an interpretation for the shift at higher frequencies of the synchrotron peak during the 2020 high state, as well as for the average harder and brighter X-ray and HE $\gamma$-ray spectra of \btwo\ compared to the low states. These trends result mainly from the average hardening, brightening and higher maximum energies of the particle spectrum injected in the regions dominating the source emission during the high state. Further contributions can include enhanced magnetic field strengths, higher bulk Lorentz factors and closer alignment with our line of sight \citep[e.g.][]{2017Natur.552..374R}. These would result in enhancing the beaming and Doppler-boosting effects.

\section{Redshift estimation}
\label{sec:redshift}

The estimation of the redshift of BL Lac objects is typically a difficult task, as the BL Lac category is characterized by definition by weak spectral lines, if any, in their optical/UV spectra.
Nonetheless, a precise measurement of the redshift is a crucial information for $\gamma$-ray blazars. 
Since $\gamma$ rays above few hundreds of GeV propagating across cosmological distances suffer from extinction by pair-production with ambient background photons composing the EBL. This leads to a redshift-dependent suppression of the observed VHE $\gamma$-ray spectrum which becomes more severe for higher energies and redshift values \citep{2013APh....43..112D}.
In addition, in BL Lac objects the photon fields external to the jet are expected to be too weak to contribute significantly to the absorption of VHE $\gamma$ rays produced by the emission regions in the jet.

The determination of the \btwo\ redshift, $z = 0.117$, dates back to \cite{1991ApJ...378...77G}. In \cite{2003A&A...400...95N}, it is suggested that the redshift of \btwo\ could actually be higher than the one quoted in literature. The argument supported by the authors derives from the analyses of imaging observations carried out with the Nordic Optical Telescope (NOT).
The optical images of a sample of BL Lac objects were fitted with a two-component spatial model accounting for the emissions from both the BL Lac core and the host galaxy.
For \btwo, an effective half-light angular radius of the host galaxy of $\theta_{\text{eff}} = 2.9 \pm 0.3\,\text{arcsec}$ was reported, which corresponds to $r_{\text{eff}} = 7.0 \pm 0.7\,\text{kpc}$ if the source is located at $z = 0.117$.
The corresponding surface brightness of the host galaxy at $\theta_{\text{eff}}$ was found equal to $\mu_{\text{eff}} = 23.26\,\pm\, 0.22\,\text{mag}\,\text{arcsec}^{-2}$. This value is incompatible with the general trend followed by BL Lac objects in the $\mu_{\text{eff}}-r_{\text{eff}}$ space, i.e. the Kormendy relation \citep{1989ARA&A..27..235K, 2009ApJS..182..216K} for elliptical galaxies, the typical host galaxies of BL Lac objects.
As the authors suggest, there is no clear reason (bad PSF, bright companions nearby...) why the 2D fit of the \btwo\ image should be significantly worse.
Therefore, by requiring that the radial profile of the source follows the Kormendy relation, an estimate of $z = 0.28 \pm 0.03$ for the source imaging redshift can be derived, based on the photometric detection of the host galaxy reported in \cite{2003A&A...400...95N}.

Given the importance of redshift estimates in evaluating properly the blazar dynamics, we applied the method presented in \cite{2010MNRAS.405L..76P} to constrain the redshift value of the source from the simultaneous spectra from MAGIC and \textit{Fermi}-LAT.
The method is based on the assumption that the slope of the EBL-corrected VHE spectrum should not be harder than the one measured by \textit{Fermi}-LAT at HE $\gamma$ rays.
An iterative procedure in fine steps of redshift from 0.01 to 2.0 was implemented. For each redshift value, the observed MAGIC spectral points were corrected for EBL extinction effects using the \cite{dominguez_ebl} model and fitted with a power-law.
The higher the redshift, the harder the photon index of the EBL-corrected VHE spectrum.
The range of redshift values at which the best-fit power-law index coincides within the statistical uncertainties with the \textit{Fermi}-LAT data yields an estimate of the upper limit $z^{\ast}$ of the source redshift.
The empirical linear relation between $z^{\ast}$ and $z_{\text{rec}}$, i.e. the estimate of the source true redshift, with the updated parameters in \cite{prandini2010b}, was employed to estimate $z_{\text{rec}}$.
This method has a reported systematic uncertainty of $0.05$ on the reconstructed redshift $z_{\text{rec}}$ \citep{2010MNRAS.405L..76P}. 
These systematic effects are mainly related to the uncertainty on the location of the intrinsic break in the spectrum of the source, as well as to a spread in the intrinsic spectral indices in the VHE band and to the uncertainties derived from the choice of different EBL models.
The application of this method to \btwo\ simultaneous \textit{Fermi}-LAT and MAGIC data in MJD $59124-59126$ yielded an estimate of $z_{\text{rec}} = 0.22 \pm 0.14_{\text{stat}} \pm 0.05_{\text{syst}}$, which is compatible with both the literature redshift $z = 0.117$ \citep{1991ApJ...378...77G} and the value that can be derived from the Kormendy relation.
The major relevance of the statistical uncertainty on this estimate reflects the poor characterization of the spectrum due to the low significance of the detection.
Nonetheless, as the inferred redshift value is coherent with the literature one, we used the latter as reference value for the broad-band SED modeling.

\section{SED modeling}
\label{sec:modeling}

The broad-band radiative emission from blazars is commonly interpreted as originated from non-thermal particles accelerated at relativistic speeds in the jet. In the simplest scenario, the emission region is assumed to be a spherical blob populated by electrons spiraling in the comoving magnetic field $B$. The blob can be imagined as an idealized emission zone which can represent a variety of relativistic plasma configurations within the jet, i.e. superluminal knots, recollimation or standing shocks. The bulk relativistic motion of the emission zone is supported by the observation of superluminal motion in radio, optical and at X rays along the jet of bright AGN, respectively at pc-scale in radio and kpc-scale in optical and X rays \citep{1971Sci...173..225W, 1971ApJ...170..207C, 1999ApJ...520..621B, 2019ApJ...879....8S}. The radiation observed in the Earth reference frame is highly beamed and Doppler-boosted by the emission zone bulk relativistic motion and by the Earth lying at small view angles, typically up to few degrees, from the jet axis.

Efficient particle acceleration occurs in blazar jets. Several mechanisms can contribute to the acceleration of particles up to the highest energies, such as magnetic reconnection \citep{2001A&A...369..694S, 2006A&A...450..887G}, first-order Fermi acceleration at relativistic shock fronts such as recollimation or internal shocks \citep{1987ApJ...315..425K, 1988MNRAS.235..997H}, stochastic (second-order Fermi) acceleration \citep[e.g.][]{2000A&A...360..789S} and shear flow acceleration \citep[e.g.][]{2002A&A...396..833R}. Several cooling mechanisms, radiative and non-radiative, contribute in parallel to the overall particle cooling \citep{1962SvA.....6..317K}. Electrons can radiatively lose energy via the emission of synchrotron photons and IC scattering of various seeds of low-energy photons.

In BL Lacs, the intensity of photon fields external to the jet, such as from the accretion disk or BLR, is thought to be rather minimal with respect to the non-thermal radiation originated from the jet, as manifested by the weakness or absence of lines in the optical/UV spectra.
SSC models, in which the electrons upscatter via IC the same synchrotron photons that they have radiated, are typically employed for describing the SEDs of BL Lacs \citep{2013ApJ...768...54B}.
In the specific context of HBLs and IBLs, SSC models are generally preferred. 
However, it has been shown in \cite{2020A&A...640A.132M} that one-zone SSC models can lead to far from energy equipartition solutions requiring extreme model parameters.
Two-zone SSC models are often successful in reproducing the broad-band SEDs of IBLs while providing more physical solutions than one-zone models, e.g. in the cases of S5\,0716+714 \citep[][]{S5_0716+714_paper} and VER\,J0521+211 \citep[][]{VERJ0521_paper}.
For a few IBLs, such as W\,Comae \citep{Acciari_2009}, 3C\,66A \citep{Abdo_2011} and PKS\,0903-57 \citep{10.1093/mnras/stab834}, one-zone solutions including IC scattering of external photon fields have been preferred over one-zone SSC solutions.
Conversely, for LBLs SSC models fail more frequently. 
Indeed, the SEDs of LBLs typically show high values of Compton dominance, i.e. the ratio between the luminosities at the peak frequencies of the high-energy and low-energy bumps \citep[e.g.][]{Finke2013}. This feature is hardly described by pure SSC models and a substantial external Compton component is often required, as in the case of OT\,081 \citep{2024MNRAS.tmp....4A}.

Alternatively, hadronic models have been employed for modeling the radiative states of several blazars \citep[e.g.][]{2017A&A...602A..25Z, 2017A&A...606A..68C}.
In recent years, hadronic models found increasing interest from the community thanks to the results from the IceCube Neutrino Observatory.
In the jets of blazars, neutrinos can be produced from the decay of mesons created in the interactions of relativistic protons with low-energy photon fields or with nuclear targets. 
IceCube reported the indication of a link between a neutrino track event and the blazar $\text{TXS}\,0506+056$ during a high state in $\gamma$ rays \citep{2018Sci...361..147I, 2018Sci...361.1378I} and the unveiling of several hot spots in the neutrino sky coincident with the directions of active galaxies \citep[][]{2020PhRvL.124e1103A, 2021ApJ...920L..45A}.
The classification of the blazar $\text{TXS}\,0506+056$ is not firmly established, although its synchrotron peak frequency would locate it into the IBL/HBL transition region \citep[$\nu_\text{synch}\lesssim10^{15}\,$Hz,][]{2018Sci...361.1378I}.
In \cite{10.1093/mnrasl/slz011}, it was pointed out that $\text{TXS}\,0506+056$ belongs to the category of masquerading BL Lacs.
These sources are characterized by luminosities as high as those observed for FSRQs, although showing synchrotron peak frequency values typical of IBLs and HBLs. In recent years, this category of blazars garnered increasing attention from the community due to their potential as neutrino source candidates \citep[e.g.][]{Fichet_de_Clairfontaine_2023, Rodriguez_LeptoHadronic}.
However, currently \btwo\ is not associated with IceCube neutrino events \citep[e.g.][]{2023ApJS..269...25A, 10.1093/mnras/staa2082}, thus we investigated a purely leptonic scenario for modeling the radiative state of the source during the 2020 flare.

As the source showed fast variability, care was given to properly select data representative of compatible source states, i.e. selecting observations that show no significant evolution in the spectrum.
Section \ref{subsec:identification_of_compatible_source_states} is devoted to the details of the data selection for the broad-band emission modeling.

\subsection{Data selection for the broad-band SED modeling}
\label{subsec:identification_of_compatible_source_states}
We divided the \btwo\ MAGIC observations into two datasets based on the results of dedicated MWL analyses in the periods surrounding the MAGIC observations reported in Table \ref{tab:magic_table}.
The first dataset is composed of the first two observations, carried out on approximately MJD 59126.9 and MJD 59127.9, i.e. October $5-6$, 2020.
The following three MAGIC observations, which constitute the second dataset, were carried out from around MJD 591230.9 to MJD 59133.9, i.e. October $9 - 11$, 2020.
In Figure \ref{fig:B2_flare_LC}, the yellow vertical band marks a region of 48 hours centered on the first two MAGIC observations times (`Period A'), extending from MJD 59126.4 to MJD 59128.4, whereas the light violet band indicates a period of 72 hours centered on the following three MAGIC observations (`Period B'), extending from MJD 59130.4 to MJD 59133.4.

\begin{figure}
\centering
\includegraphics[width = 0.95\columnwidth]{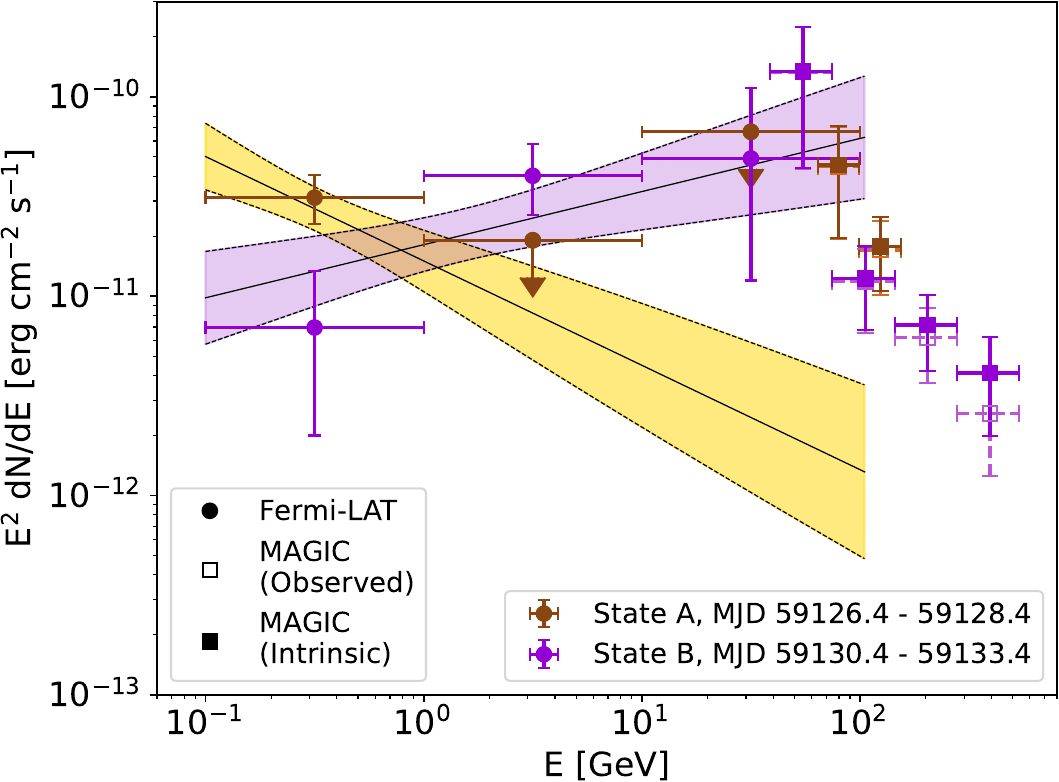}
\caption{
Spectral energy distribution at HE and VHE $\gamma$ rays of B2 1811+31 during periods A and B. For the two periods, spectral points are shown in brown and violet and the $1\sigma$ uncertainty shaded bands are in yellow and light violet, respectively.
The points with full round markers are derived from \textit{Fermi}-LAT data.
For the two periods, the best-fit parameters of the spectrum reconstructed from $\textit{Fermi}$-LAT observations are reported in Table \ref{tab:summ_table_spectral_B2_stateA_B}. The points with full square markers show the MAGIC EBL-corrected SEDs. The observed MAGIC spectral points are indicated with hollow square markers and dashed error bars.}
\label{fig:B2_statesAB_Fermi_LAT}
\end{figure}

In order to look for differences in the source states during the two periods, dedicated \textit{Fermi}-LAT analyses were carried out using the same prescriptions presented in Section \ref{subsec:Fermi_analysis}. 
The results are summarized in Table \ref{tab:summ_table_spectral_B2_stateA_B} and shown in Figure \ref{fig:B2_statesAB_Fermi_LAT}.
The TS for the detection of the source in the two periods were TS$_{\text{A}} = 24$ and TS$_{\text{B}} = 37$.
The reconstructed HE $\gamma$-ray spectra during the A and B periods are significantly different, providing evidence for different source states in the two periods.
The HE $\gamma$-ray spectrum is harder in state B than in state A, the resulting PL spectral indices being $\Gamma_{\text{A}} = 2.5\,\pm\,0.3$ and $\Gamma_{\text{B}} = 1.7\,\pm\,0.2$. The reconstructed fluxes in the energy range above $100\,$MeV during the two states are different as well, being $21\,\pm\,7$ and $8\,\pm\,4$, in units of 10$^{-8}$ ph cm$^{-2}$ s$^{-1}$.
We also performed analyses of \textit{Fermi}-LAT data over periods of 24 hours centered on each MAGIC observation night.
Although these short-duration analyses employ datasets with poorer statistics with respect to analyses on longer integration times, they followed the general trend of showing harder spectra in period B than in A.
It is worth pointing out that these results follow the hint of softer-when-brighter trend within the 2020 high state described in Section \ref{subsec:intraband_correlations_HEgamma_rays}.

In Figure \ref{fig:B2_statesAB_Fermi_LAT}, the EBL-corrected MAGIC spectra derived from the observations in periods A and B are shown (Section \ref{subsec:MAGIC_analysis}). While in period B there is a smooth connection between the \textit{Fermi}-LAT and MAGIC spectra, in period A the two spectra do not seem to be easily compatible.
As reported in Section \ref{sec:variability_cross-correlations}, during the high state the source shows variability in timescales of few hours. Thus, this effect might be due to fast variations in the $\gamma$-ray spectrum of the source that may have been observed by MAGIC and that were smoothed out in the analysis of \textit{Fermi}-LAT data integrated over 24 and 48 hours.
The \btwo\ SED in state B was constructed using data simultaneous or quasi-simultaneous with the MAGIC observations.
We merged the data from the three MAGIC observations within state B and included in the SED to be modeled the resulting average EBL-corrected source spectrum presented in Section \ref{subsec:MAGIC_analysis}.

We did not perform the modeling of the source state in period A as it is poorly constrained in both the HE and VHE $\gamma$-ray bands. Therefore we performed the modeling of a single snapshot of the source broad-band SED in flaring state, in particular the one in period B, rather than a time-dependent modeling.
    
\subsection{Broad-band SED modeling}
We employed the {\fontfamily{cmtt}\selectfont agnpy} \citep{2022A&A...660A..18N} open-source code for computing the synchrotron and SSC SEDs for modeling the emission of the source during the high state. The models were investigated using a `fit-by-eye' strategy.
As discussed previously, the broad-band SED of BL Lacs is generally modelled with SSC models due to the low intensity of photon fields external to the jet.

\begin{figure*} [t]
\centering
 \subfloat[One-zone SSC modeling]{\includegraphics[width=1\columnwidth]{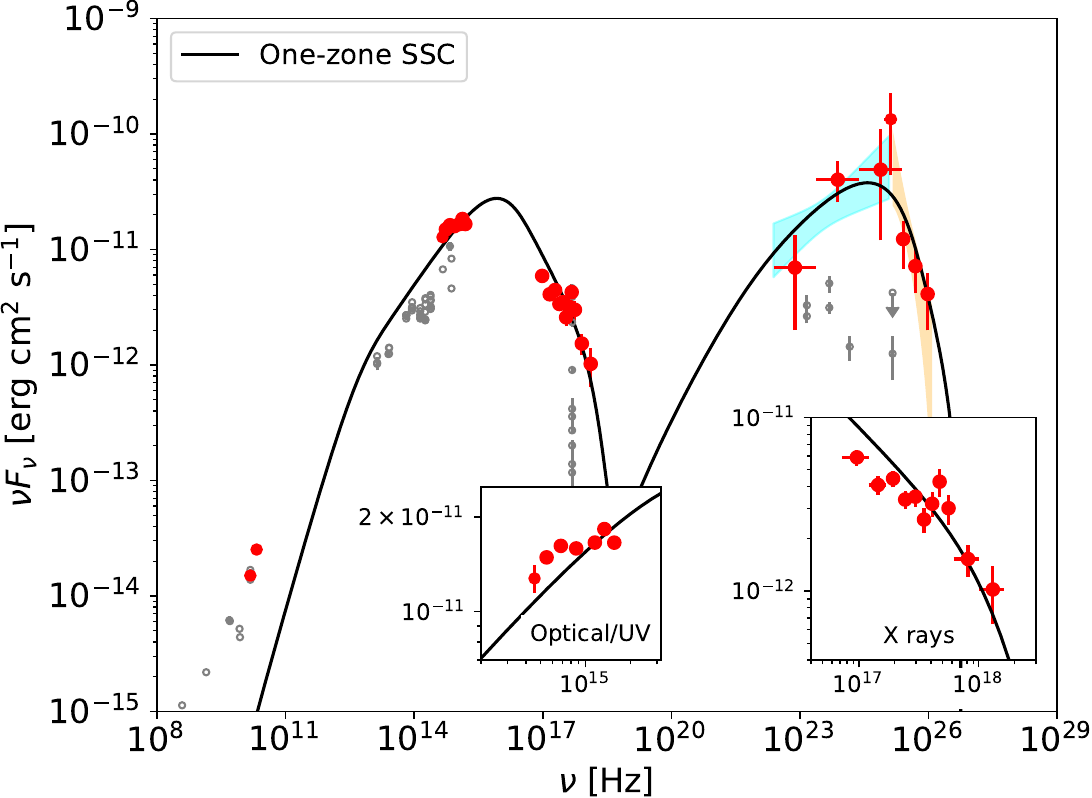} \label{fig:B2_modeling_one_zone}}
 \hfill
 \subfloat[Two-zone SSC modeling]{\includegraphics[width=1\columnwidth]{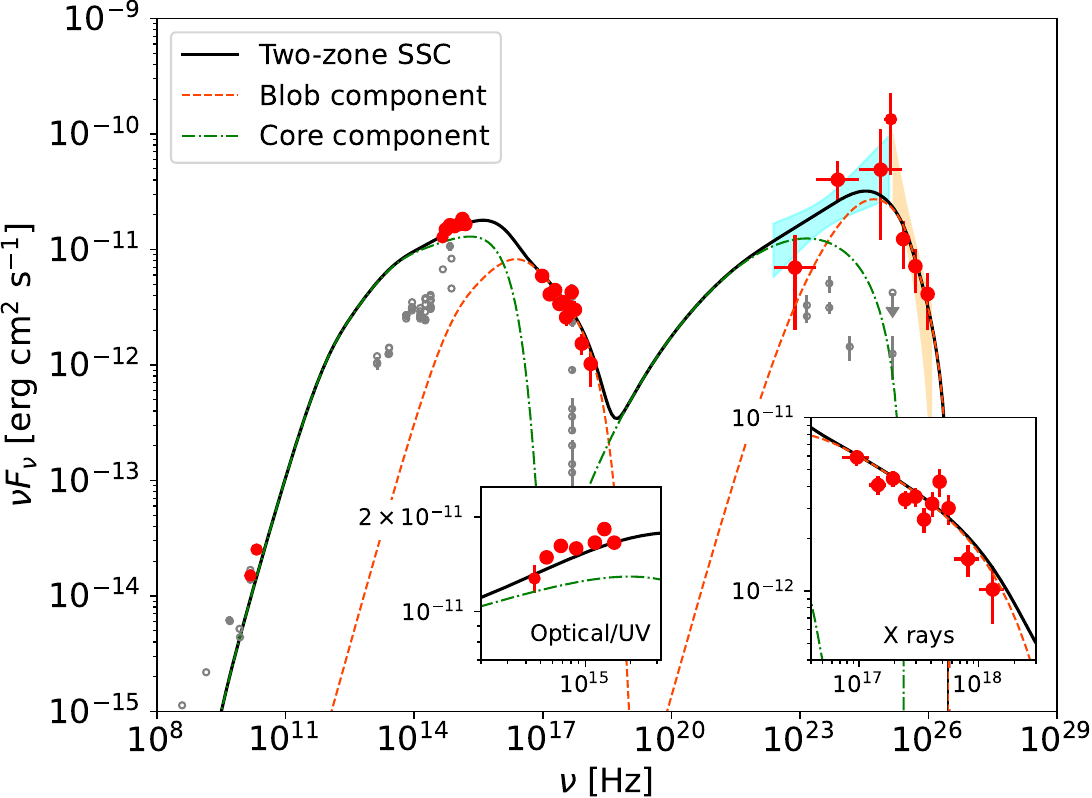}
 \label{fig:B2_modeling_two_zones} } 
 \\
 \caption{
 Leptonic modeling of the broad-band SED of \btwo\ during Period B within the 2020 high state in VHE $\gamma$ rays, marked by the red points. Archival data are indicated by the hollow gray points.
 The blue and orange shaded areas indicate the $1\sigma$ uncertainty bands on the intrinsic spectrum of the source derived from \textit{Fermi}-LAT and MAGIC observations, respectively.
 The two insets show a zoom-in on the optical/UV and X-ray bands, respectively. For the two-zone SSC model, the contributions from the blob and core are indicated respectively by the orange and green lines, whereas the total is computed as the sum of the two.
 }
 \label{fig:B2_modeling}
\end{figure*}

\begin{table*} [t]
\small
\centering
\caption{SED modeling parameters for the one-zone and two-zone SSC model of the 2020 VHE $\gamma$-ray high state of B2 1811+31.
}
\label{tab:B2_modeling_results}
\begin{tabular}{ccccccccccc}
\hline
Model & $\gamma_\textrm{min}$ & $\gamma_\textrm{b}$ & $\gamma_\textrm{max}$ & $p_1$ & $p_2$ & $B$ & $K_{e}$ & $R_\text{b}$ & $\delta_{\text{D}}$ & $u_e/u_B$\\
(region) & $[\times 10^3]$ & $[\times 10^4]$ & $[\times 10^5]$ & & & [G] & $[\textrm{cm}^{-3}]$ & $[\times 10^{15} \textrm{cm}]$ & &\\
\hline
One-zone & 0.7 & 3.2 & 3.0 & 1.9 & 4.4 & 0.13 & 7.6 & 10.6 & 20 & 31.3 \\
\hline
Two-zone (blob) & 6.0 & 4.0 & 3.0 & 2.0 & 3.8 & 0.38 & 7.95 & 4.0 & 10 & 18.3 \\
Two-zone (core) & 0.5 & 0.38 & 0.7 & 1.8 & 2.7 & 0.17 & 0.74 & 210 & 4 & 1.0 \\
\hline
\end{tabular}
\tablefoot{
Columns:
(1) Model.
(2), (3), (4) Minimum, break and maximum electron Lorentz factors.
(5), (6) Slopes of the electron spectrum around the break.
(7) Magnetic field strength.
(8) Electron density.
(9) Radius of the emission region.
(10) Doppler factor.
(11) Equipartition parameter.
}
\end{table*}

\subsubsection{One-zone modeling of the SED}
\label{subsubsec:one-zone_modeling}

We first adopted a one-zone SSC scenario, with a unique spherical emission region being responsible for most of the radiative output.
The emission region in blazars is moving with relativistic bulk Lorentz factor $\Gamma$ along an axis forming a viewing angle $\theta$ with the line of sight from Earth.
It is assumed to be populated by relativistic electrons with electron energy density (EED) $n_{\text{e}}\left( \gamma \right)$, with $\gamma$ being the Lorentz factors in the frame comoving with the emission region. 
We employed a broken power-law EED defined as
\begin{equation}
\label{BPL_EED}
    n_{\text{e}}\left( \gamma \right) = 
    k_\text{e} \left[ 
    \left( \frac{\gamma}{\gamma_\text{b}} \right)^{-p_1}
    H \left( \gamma; \gamma_\text{min}, \gamma_\text{b} \right)
    + \left( \frac{\gamma}{\gamma_\text{b}} \right)^{-p_2}
    H \left( \gamma; \gamma_\text{b}, \gamma_\text{max} \right)
    \right]
\end{equation}
between $\gamma_{\text{min}}$ and $\gamma_{\text{max}}$, with break at $\gamma_{\text{b}}$, spectral indices $p_1$ and $p_2$ below and above the break, respectively.
The $H$ function is the Heaviside function, defined such that $H\left( x; a, b \right) = 1$ if $a \leq x \leq b$ and $H\left( x; a, b \right) = 0$ otherwise.
$k_\text{e}$ is the EED value at $\gamma_\text{b}$, in cm$^{-3}$, and the electron density is defined as $K_\text{e} = \int_{ \gamma_{\text{min}} }^{\gamma_{\text{max}}} d\gamma \, n_{e} \left( \gamma \right)$.
The magnetic field tangled to the region is $B$.
The energy densities in electrons and magnetic field of the emission region are $u_\text{e} = m_{\text{e}} c^{2} \int_{ \gamma_{\text{min}} }^{\gamma_{\text{max}}} d\gamma \, \gamma \, n_{e} \left( \gamma \right)$ and $u_{B} = \frac{B^{2}}{8 \pi}$ in Gaussian cgs units, respectively.
The energy equipartition parameter $k_{\text{eq}}$ is defined by $k_{\text{eq}}=\frac{u_{e}}{u_{B}}$.

The MWL spectral analyses presented in Section \ref{sec:analysis} can be employed to estimate the EED spectral indices $p_1$ and $p_2$.
Synchrotron radiation theory \citep[e.g.][]{2010herb.book.....D} predicts that a population of electrons with EED $n_{\text{e}}\left( \gamma \right) \propto \gamma^{-p}$ radiates a synchrotron SED $\nu F_{\nu, \text{synch}} \propto \nu^{\frac{3-p}{2}}$.
As shown in Section \ref{sec:source_classification}, the X-ray spectrum of \btwo\ lies deep in the falling edge of the synchrotron bump and it is produced by the electrons with $\gamma \gg \gamma_\text{b}$, where $n_{\text{e}}\left( \gamma \right) \propto \gamma^{-p_2}$.
In Period B, the X-ray flux is well described with a PL (Eq. \ref{eq:PL}) with spectral index $\Gamma_{\text{X}} = 2.58 \pm 0.10$ (Table \ref{tab:results_XRT_B2}). Hence, we derive $p_{2} = 2 \Gamma_{\text{X}} - 1 \approx 4.2 \pm 0.2$.
Conversely, the \textit{Swift}-UVOT 2020 high-state data yield that $\nu F_{\nu} \propto \nu^{ 0.25 \pm 0.10 }$ in the optical/UV band. This would imply a spectral index of $2.5 \pm 0.2$ for the electron population dominating the radiative output in this band. However, as shown in Section \ref{sec:source_classification}, the high-state synchrotron peak lies close to the optical/UV band, hence the electrons contributing to the flux in this band have Lorentz factors close to $\gamma_\text{b}$, where the EED spectral index changes. Thus, for the \btwo\ high state, the argument to estimate $p_1$ from the optical/UV spectral index cannot be applied safely.

The Doppler factor (Eq. \ref{eq:Doppler_factor}) is a crucial parameter for the SSC models, since it strongly affects the properties of the observed radiation ($\nu F_{\nu} \propto \delta_{\text{D}}^4$).
In literature, standard values commonly used for blazars single-zone models are $\delta_{\text{D}} \approx 10-20$ \citep[e.g.][]{2009A&A...494..527H}.
In the construction of the one-zone model solution, we tried several values for the Doppler factor $\delta_\text{D}$, up to values around 30.
While keeping fixed the $\delta_\text{D}$ value, the other model parameters were tuned to account for the observed SED and for the constraint deriving from the observed variability at HE $\gamma$ rays, presented in Section \ref{subsec:variability}.
A single-zone SSC solution with $\delta_\text{D} = 20$ is shown in the left panel of Figure \ref{fig:B2_modeling_one_zone}.
The parameters of this solution are reported in Table \ref{tab:B2_modeling_results}.
In Section \ref{subsec:variability}, we derived an upper limit for the emission region size of $\text{R}_{\text{max}} \approx \left( 6 - 12 \right) \times 10^{15}\,\text{cm}$ for $\delta_{\text{D}} = 20$.
The solution in Figure \ref{fig:B2_modeling_one_zone} has $\delta_{\text{D}} = 20$, $R_\text{b} = 10.6 \times 10^{15}\,\textrm{cm}$ and magnetic field $B = 0.13\,\text{G}$.
In the one-zone solutions obtained with $\delta_\text{D}$ fixed to values lower than 20, higher values of both the region size and the magnetic field were required to reproduce the spectra in all the energy bands.
These solutions violated the constraint imposed by the observed variability at HE $\gamma$ rays and showed a worse agreement with the observed SED with respect to the solution with $\delta_\text{D} = 20$ shown in Figure \ref{fig:B2_modeling_one_zone}.

As discussed in \cite{2016MNRAS.456.2374T}, the magnetic field strengths for single-zone models tend to be significantly lower than the values required for equipartition.
The one-zone SSC scenario presented in Figure \ref{fig:B2_modeling_one_zone} has $k_{\text{eq}} = 31.3$. In literature, values of the order of 100 and higher are usually found for single-zone models \citep[e.g.][]{2020A&A...640A.132M}. Although the reported solution has lower $k_{\text{eq}}$, it is still far from equipartition ($k_{\text{eq}}=1$).
To reproduce the optical-to-X-ray spectrum, the break between the spectral indices $p_1$ and $p_2$ is required to be quite extreme, with $p_1 = 2.0$ and $p_2 = 4.4$.
The resulting $p_2$ value is compatible within the uncertainties with the estimate $p_2 \approx 4.2 \pm 0.2$ derived from the X-ray spectral index.
Such an extreme EED spectral break can be a consequence of describing the broad-band SED with a model composed by a single emission region.
The presented one-zone SSC solution underestimates the radio flux. Single zones employed for modeling high-state SEDs are known to hardly manage to account for the radio emission \citep[e.g.][]{2020A&A...640A.132M}.
This is due to the dominance of synchrotron-self-absorption effects at radio frequencies for compact and energetic emission regions.

\subsubsection{Two-zone modeling of the SED}
\label{subsubsec:two-zone_modeling}

We investigated a two-zone SSC model in line with the one presented in \cite{2011A&A...534A..86T}, with a few modifications. Due to the BL Lac nature of the source, we neglect the IC scattering contribution from photon seeds external to the jet. Given the MWL properties presented in the previous sections, we propose a physical scenario as follows. A small and energetic region, denoted as `blob', is required to dominate the emissions in the X-ray and VHE $\gamma$-ray bands. The second zone, denoted as `core', is required to be larger and less energetic (lower bulk Lorentz factor and EED at lower $\gamma$ values) than the blob. This larger region dominates the optical/UV emission and yields a significant contribution to the HE $\gamma$-ray band, as suggested by the observed correlation between the long-term fluxes in these energy bands, presented in Section \ref{subsec:cross-correlations}. The blob region is considered to be located closer to the central engine than the core region, which would account for the blob region being more energetic than the core. Moreover, as the magnetic field strength along the relativistic jet scales with the distance from the central engine, the magnetic field tangled to the core is required to be smaller than the blob one. As the core region is assumed to be located further along the jet than the blob one, the interplay between the two regions is expected to be negligible, i.e. the synchrotron photons of one zone do not serve significantly as seed photons for the IC of the electrons from the other region \citep[][]{2011A&A...534A..86T}. The construction of the two-zone model solution was performed following a similar procedure to that employed for the one-zone model solution.
Several values for the Doppler factors $\delta_\text{D}$ of two regions were tried. For each value of $\delta_\text{D}$, the parameter space was explored accounting for the constraints resulting from the MWL analyses presented in the Section \ref{sec:variability_cross-correlations}.
As described later in this section, in the parameter space scan, we also investigated the possibility to account for the observed broad-band SED in a self-consistent scenario, i.e. with emission regions in which the particle distributions are compatible with those expected at equilibrium under the effects of escape from the emission region and radiative cooling via the emission of synchrotron and SSC radiation \citep[][]{1996ApJ...463..555I, 2020Galax...8...72C}.
The resulting two-zone SSC solution is presented in Figure \ref{fig:B2_modeling_two_zones}, the parameters are reported in Table \ref{tab:B2_modeling_results}.
This solution characterizes the optical-to-X-ray SED more naturally than the single-zone one (Appendix \ref{sec:appendix_chi_square}), without invoking hard spectral breaks in the distributions of the electrons in both regions.
The $p_{1,\,\text{core}}$ and $p_{2,\, \text{core}}$ parameters of the core region are $1.8$ and $2.7$, respectively, with a spectral break close to 1.
The spectral break of the blob region EED also mitigates, being $p_{1,\,\text{blob}}=2.0$ and $p_{2,\,\text{blob}}=3.8$. As discussed above, the $p_2$ parameter of the region dominating the X-ray flux is required to be soft by the observed X-ray spectral index.
The relativistic Doppler factors of the two regions are $\delta_{\text{D, blob}} = 10$ and $\delta_{\text{D, core}} = 4$.
For $\delta_{\text{D}} = 10$, the variability analysis reported in Section \ref{subsec:variability} yields an upper limit of $\text{R}_{\text{max}} \approx \left( 3 - 6 \right) \times 10^{15} \,\text{cm}$ on the size of the zone dominating the $\gamma$-ray flux during the high state. The size of the blob region in the presented solution is $R_{\text{blob}} = 4.1 \times 10^{15} \, \text{cm}$, thus it satisfies the constraint set by the causality argument.

In addition, $\delta_{\text{D, blob}}=10$ is consistent with the range of Doppler factor values allowed by the constraint for the region to be transparent to $\gamma$ rays up to the highest-energy ones observed by MAGIC, i.e. $E_\text{ph, max} = 400\,\text{GeV}$ in the reference frame of the observer (Figure \ref{fig:B2_statesAB_Fermi_LAT}). 
The blob region dominates the source emission in the X-ray and VHE $\gamma$-ray bands. VHE $\gamma$-ray photons can interact with soft X rays to produce $e^{+}e^{-}$ pairs, hence suppressing part of the VHE $\gamma$-ray flux emitted by the region \citep[e.g.][]{2010herb.book.....D}.
A lower limit to the Doppler factor of the blob can be derived by requiring that the emission region is not opaque to the observed highest-energy $\gamma$ rays. This lower limit can be estimated as \cite[e.g.][]{1995MNRAS.273..583D}
\begin{equation*}
    \delta_{\text{D, blob}} > \sqrt[6]{\frac{\sigma_\text{T} d_\text{L}^2 \left(1+z\right)^2 f_\text{X} \epsilon_\text{max}}{4 m_\text{e} c^4 t_\text{var}}},
\end{equation*}
being $\sigma_\text{T}$ the Thomson cross section, $d_\text{L}$ the luminosity distance, $f_\text{X}$ the observed X-ray flux, $t_\text{var}$ the fast variability timescale and $\epsilon_\text{max} = \frac{E_\text{ph, max}}{m_\text{e} c^2}$.
Using $z=0.117$ and $f_\text{X}\approx 10^{-11} \, \text{erg}\,\text{cm}^{-2}\text{s}^{-1}$ (Table \ref{tab:results_XRT_B2}, Figure \ref{fig:B2_modeling_two_zones}), we obtain for $t_\text{var}=3-6\,\text{h}$ (Table \ref{tab:variability_table_B2}) a lower limit of $8-9$ to $\delta_{\text{D, blob}}$. 
Hence, the value $\delta_{\text{D, blob}}=10$ would allow the blob region to be significantly transparent to $\gamma$ rays up to the ones at the highest energies observed by MAGIC.

As far as the variability analyses presented in Section \ref{subsec:variability_weekly}, it is interesting to notice that the light-crossing time argument on $t_\textrm{rise}= \left( 18 \pm 4 \right)\,$d, observed in correspondence to the VHE detection by MAGIC, yields an estimated upper limit on the emission region size $R_{\text{max}} = \left( 1.8 \pm 0.4 \right) \times 10^{17}$ cm for a Doppler factor $\delta_\text{D} = 4$, using Eq. \ref{eq:R_max}.
This value is fully compatible with the value $R_{\textrm{core}} = 2.1 \times 10^{17}$ cm employed for the core region.
The value of $t_\textrm{rise}= \left( 18 \pm 4 \right)\,$d was estimated from the variations in the HE $\gamma$-ray flux using the \textit{Fermi}-LAT weekly-binned light-curve. 
In the two-zone SSC solution shown in Figure \ref{fig:B2_modeling_two_zones}, both the core and blob regions contribute significantly to the HE $\gamma$-ray total flux. However, the two regions produce variability on very different timescales, i.e. $\sim 10\,$days for the core and $\sim 5\,$ hours for the blob. Therefore, the core is likely responsible for the slower variations in the HE $\gamma$-ray flux at the timescales from days to weeks, shown in Figure \ref{fig:b2_LAT_weekly_optical}, whereas the blob is likely producing the short-timescale variations on top of the long-term ones, shown in Figure \ref{fig:B2_flare_LC}.

To account properly for both the X-ray and VHE $\gamma$-ray spectra, the energy range of the electrons populating the blob region extends up to $E_\text{max} = m_\text{e} \gamma_{\text{max}} c^2 \approx 150\,\text{GeV}$.
Moreover, the core component is able to account more naturally for the radio flux, which was severely underestimated by the one-zone solution.

In the presented solution, the core region is at energy equipartition, whereas the blob equipartition coefficient is improved ($k_{\text{eq}} \approx 18$) with respect to the one-zone scenario. 
This is consistent with \cite{2020A&A...640A.132M}, in which the authors presented a systematic study on two-component scenarios employed for modeling blazar broad-band SEDs and found that two-zone models are systematically closer to equipartition with respect to single-zone ones, but still particle-dominated.

\begin{figure}
\centering
\includegraphics[width = 1.\columnwidth]{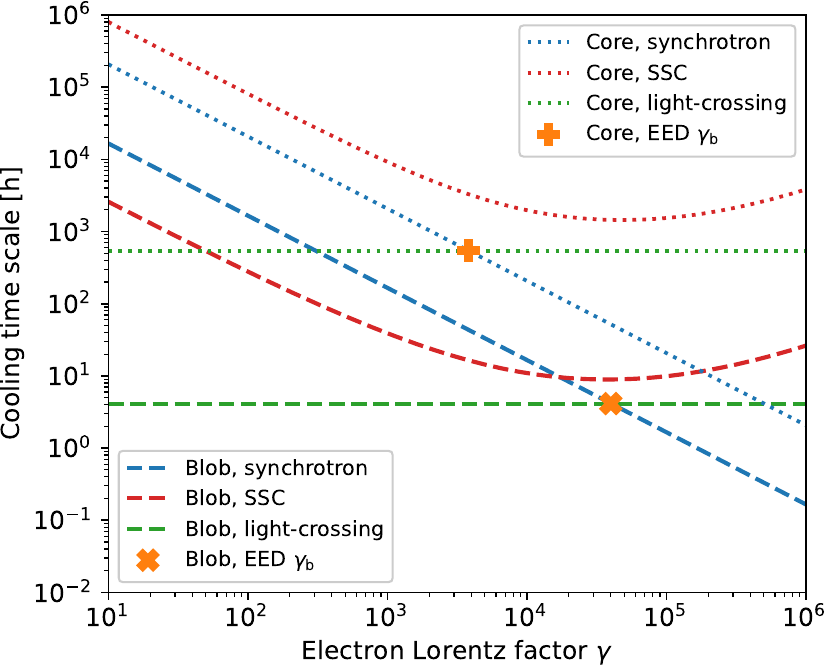}
\caption{
Competition among the cooling timescales for the two-zone model in Figure \ref{fig:B2_modeling_two_zones}, as a function of the electron Lorentz factor $\gamma$. 
The times are in the observer frame, i.e. corrected by the corresponding $\frac{1+z}{\delta_\text{D}}$ with respect to the frames comoving with the two regions. Dashed (dotted) curves refer to the blob (core) region. Blue and red curves indicate the synchrotron and SSC cooling timescales, respectively, whereas the green lines represent the light-crossing times.
The orange markers indicate the $\gamma_\text{b}$ values of the EED within the two emission regions (Table \ref{tab:B2_modeling_results}).
}
\label{fig:B2_cooling_timescales}
\end{figure}

As previously mentioned, in the construction of the one-zone and two-zone model solutions, we investigated the possibility to account for the observed broad-band SED with self-consistent particle populations in the emission regions.
The temporal evolution of the EED is typically studied through a continuity equation under the effects of particle acceleration/injection, escape, adiabatic expansion and radiative cooling \citep[e.g.][]{1962SvA.....6..317K}.
Figure \ref{fig:B2_cooling_timescales} shows, for the two regions presented in Figure \ref{fig:B2_modeling_two_zones}, the competition among the light-crossing timescale $t_{\text{lc}}$ and the main radiative cooling timescales.
The light-crossing timescale can be interpreted as the characteristic time for the particle escape from the emission region.
In the two-zone solution in Table \ref{tab:B2_modeling_results}, all particles are ultra-relativistic, hence the light-crossing time is energy-independent, with $t_{\text{lc}} \sim R_\text{b}/c$.
The timescales for synchrotron energy losses were computed as \citep[][]{2010herb.book.....D}
\begin{equation}
 t_{\text{synch}} \left( \gamma \right) = \gamma \left( \frac{d \gamma}{dt} \right)_\text{synch}^{-1} = \frac{3 m_{e} c}{4 \sigma_{T} u_{B} \gamma} \propto \gamma^{-1} \, .
 \label{eq:synchrotron_cooling_timescales}
\end{equation}

For each emission region, the SSC cooling timescale corresponds to that of the IC scattering off the (isotropic) synchrotron photon number density $n_\text{ph} (\epsilon) = \frac{dN}{dV d\epsilon}$, being $\epsilon = \frac{h \nu}{m_\text{e} c^2}$ the dimensionless photon energy and $n_\text{ph} (\epsilon) \, dV \, d\epsilon$ the number of photons in a volume $dV$ having dimensionless energy between $\epsilon$ and $\epsilon + d\epsilon$. Following \cite{2022A&A...660A..18N}, $n_\text{ph} (\epsilon)$ of the synchrotron fields in the two regions were computed as 
\begin{equation}
    \label{eq:n_ph}
    n_\text{ph}(\epsilon) = \frac{1}{m_\text{e} c^2 \epsilon} u_\text{ph}(\epsilon) = \frac{3}{4} \frac{1}{m_\text{e} c^2 \epsilon} \frac{R_\text{b}}{c} \frac{ \sqrt{3} e^3 B }{h} \int_1^{\infty} d \gamma \, n_\text{e}(\gamma) \, R(x) \, ,
\end{equation}
where $u_\text{ph}(\epsilon)$ is the photon energy density, the factor 3/4 accounts for averaging the radiation in a sphere and $R(x)$ is the single-electron synchrotron spectral power averaged over the pitch angles \citep[e.g. Eq. 7.45 in][]{2010herb.book.....D}, expressed as a function of the variable $x=\frac{4 \pi \epsilon m_\text{e}^2 c^3}{3 e B h \gamma}$.
We computed the cooling timescales $t_{\text{SSC}} \left( \gamma \right) = \gamma \left( \frac{d \gamma }{dt} \right)_\text{SSC}^{-1}$ of relativistic electrons due to IC scattering with the isotropic synchrotron radiation field as in \cite{2005MNRAS.363..954M}, appendix C.
Therefore we accounted for IC scattering energy losses both in Thomson and Klein-Nishina (KN) regimes, in which $t_\text{SSC} \propto \gamma^{-1}$ and $t_\text{SSC} \propto \gamma/\log(\gamma)$, respectively. The steady-state solution of the EED continuity equation when synchrotron and IC scattering (Thomson limit only) are the main radiative cooling mechanisms have EED spectral break $p_2 - p_1 = 1$ and $\gamma_\text{b}$ equal to the value at which the escape and radiative cooling timescales coincide
\citep[e.g.][]{1996ApJ...463..555I, 2020Galax...8...72C}.
We find that, for both regions in Figure \ref{fig:B2_modeling_two_zones}, the $\gamma_\text{b}$ values in our solution (Table \ref{tab:B2_modeling_results}) appear to be equal those obtained comparing the radiative cooling timescales with $t_{\text{lc}}$ (Figure \ref{fig:B2_cooling_timescales}).
The spectral break in the core region is $p_\text{2, core} - p_\text{1, core} \approx 0.9$, thus it is consistent with the one expected at equilibrium when $t_\text{cool} \propto \gamma^{-1}$. Conversely, for the blob of the two-zone solution, a stronger break is necessary to account for the observed spectra. This suggests that further effects may be relevant, such as the particle distribution being hardly at equilibrium or possible effects related to cooling in the KN regime. In contrast, in the investigation of the one-zone model presented in Section \ref{subsubsec:one-zone_modeling}, we find it is not possible to properly account for the observed SED in all energy bands while requiring that the $\gamma_{\text{break}}$ is the one expected at equilibrium or requiring that the spectral break is equal to 1. Hence, in Figure \ref{fig:B2_modeling_one_zone}, we present the solution which fits better the broad-band SED, although the value of $\gamma_{\text{break}}$ for which the synchrotron and light-crossing timescales coincide is $\gamma_{\text{break}}=1.3 \times 10^{5}$. Although the two-zone model solution presented here is not time-dependent, this analysis supports the self-consistency of the proposed physical two-zone model scenario for the MWL emissions of \btwo\ during the 2020 high state.

In SSC models, the predicted $\nu F_\nu$ from an emission region depends on the bulk Lorentz factor $\Gamma$ and viewing angle $\theta$ only through $\delta_\text{D}$, i.e. the predicted flux does not depend independently on $\Gamma$ and $\theta$.
For an emission region with given $\delta_\text{D}$, Eq. \ref{eq:Doppler_factor} implies that $\Gamma \geq \Gamma_\text{min} = \frac{1 + \delta_\text{D}^2}{2 \delta_\text{D}}$ and $\theta \leq \theta_\text{max} = \arccos{\sqrt{1 - \delta_\text{D}^{-2}}}$, being $\theta_\text{max}$ corresponding to $\Gamma=\delta_{D}$. In the presented two-zone solution, the blob region has $\delta_\text{D} = 10$, which results in an upper limit of $\theta_\text{max} \approx 5.7^{\circ}$ to the jet viewing angle. Conversely, the jet power depends only on $\Gamma$ and is independent of $\theta$. For each emission region, from the energy densities $u_\text{e}$ and $u_\text{B}$ in electrons and magnetic field, the corresponding luminosities can be computed as $L_\text{e} = 2 \pi R_{\text{b}}^{2} \, \beta c \Gamma^{2} u_\text{e}$ and $L_\text{B} = 2 \pi R_{\text{b}}^{2} \, \beta c \Gamma^{2} u_\text{B}$, respectively. Thus, for viewing angles between $1^{\circ}$ and $5^{\circ}$ and for $\Gamma < \delta_\text{D}$, the jet luminosity in particles of the blob (core) region is $L_\text{e}\approx 1 \times 10^{43} \, \text{erg s}^{-1}$ ($4 \times 10^{43} \, \text{erg s}^{-1}$), whereas the corresponding jet luminosity in magnetic field is scaled down by a factor $k_\text{eq} = 18.3$ ($k_\text{eq} = 1$).

Long-term MWL campaigns for several blazars have shown radio emission correlated with a significant delay, from weeks to years, with respect to the optical and $\gamma$-ray emission \citep[e.g.][]{2010ApJ...722L...7P}. This evidence is interpreted as related to synchrotron-self-absorption effects and to the emission region becoming transparent in radio later than at higher frequencies \citep[e.g.][]{1979ApJ...232...34B}. In the case of \btwo, the 15 GHz long-term light-curve shown in the bottom panel of Figure \ref{fig:B2_long_term_LC} reports a high state in 2021, nearly a year after the 2020 VHE outburst.  However, the cross-correlation analysis presented in Section \ref{subsec:cross-correlations} showed that the radio flux is not significantly correlated with the emissions in the optical and HE $\gamma$-ray bands, even when delays up to few years are taken into account. This suggests that other regions, possibly located further along the jet, may dominate the radio flux from the source. Because of this, in the two-component solution shown in Figure \ref{fig:B2_modeling_two_zones}, we left the core region able to account for most of the radio emission (up to $\sim 50 \%$), but not entirely. For instance, the core would account entirely for the observed radio data if $\gamma_\text{min}$ was lowered from $0.5 \times 10^3$ to $0.35 \times 10^3$ and the electron density $K_\text{e}$ was increased to $1.0\,\text{cm}^{-3}$, keeping the other parameters in Table \ref{tab:B2_modeling_results} fixed. With such modifications, the agreement between the data and the flux predicted from the two-zone model at higher frequencies would not be compromised and the energy equipartition coefficient of the core region would be $k_\text{eq} \approx 1.1$. One possibility for the interpretation of the 2021 radio flare is adiabatic expansion of the core region during its propagation along the jet \citep{2022A&A...658A.173T}. However, we consider the available data not sufficient to confirm nor exclude that the origin of the 2021 radio flare is related to the 2020 MWL outburst.

The parameter space in SSC models is quite vast and SSC solutions are usually degenerate. The solution shown in Fig \ref{fig:B2_modeling_two_zones} represents one possible set of parameters, but all observational constraints have been taken into account, the model well reproduces the observed SED and the consistency of the proposed physical scenario for the two regions has been investigated.

\section{Summary and conclusions}
\label{sec:conclusions}

In this paper, we presented a long-term MWL view of \btwo, with the main emphasis given to the 2020 $\gamma$-ray flare of the source. During this high-state period, the MAGIC telescopes reported the the first-time detection of VHE $\gamma$-ray emission from the source. The MWL observational campaign during the flaring state was organized after the detection of a high state in HE $\gamma$ rays by the $\textit{Fermi}$-LAT and it provided coverage from radio to the VHE band.

The long-term MWL behavior of \btwo\ indicates that the optical and $\gamma$-ray emissions correlate, whereas the evolution of the radio flux is significantly different from that at higher frequencies. This provides evidence for the same emission regions dominating the radiative output in the optical and HE $\gamma$-ray bands. The 2020 flaring state occurred at the apex of an increasing trend in the optical band that persisted for approximately eight years. In addition, after the flare, the source showed a falling trend of the optical flux that led the source to a state compatible to the one before the rising trend. The duration of the falling trend was approximately two years, much shorter than the rising trend. Indications of this long-term rising-peaking-falling evolution can be found in X rays and HE $\gamma$ rays as well. The long-term X-ray flux around 1 keV exhibits the harder-when-brighter trend and it shows variations by more than 2 orders of magnitude.

We confirmed the IBL classification of the source during the low state making use of archival IR data and of dedicated optical/UV and X-ray analyses. During the 2020 high state, the average HE $\gamma$-ray spectrum became harder compared to the low states. A similar behaviour has been observed in X rays. Conversely, during different activity periods, we find harder-when-brighter trends in X rays and a hint of softer-when-brighter trends at HE $\gamma$ rays. In addition, we claim a significant shift of the synchrotron peak frequency during the 2020 flare, which led the source to a borderline IBL/HBL state.

Analysis of the fast variability in the HE  $\gamma$-ray band during the 2020 flare was used to estimate an upper limit to the size of the emission region dominating the $\gamma$-ray output. We tested one-zone and two-zone SSC scenarios for modeling the broad-band SED probed with the MAGIC telescopes. The widely used one-zone SSC model provides a good estimation of the broad-band emission, with some known limitations, such as the underestimation of the radio flux and a hard break in the electrons distribution. In addition, this scenario is far from energy equipartition. For the two-zone model, we investigated a scenario with a small and fast blob dominating the flux at X rays and VHE $\gamma$ rays, and a larger and slower blob, denoted as core, that dominates the optical flux and contributes significantly at HE $\gamma$ rays.
The two-zone model manages to describe the SED in the optical/UV and X-ray bands without invoking extreme parameters for the electrons distribution.
The proposed two-zone SSC solution is closer to energy equipartition and accounts more naturally for the radio flux than the one-zone scenario.
In addition, the spectral breaks of the EEDs within the two emission regions are comparable to the ones expected from radiative cooling, supporting self-consistency for the proposed model.

\begin{acknowledgements}
      We would like to thank the Instituto de Astrof\'{\i}sica de Canarias for the excellent working conditions at the Observatorio del Roque de los Muchachos in La Palma. The financial support of the German BMBF, MPG and HGF; the Italian INFN and INAF; the Swiss National Fund SNF; the grants PID2019-104114RB-C31, PID2019-104114RB-C32, PID2019-104114RB-C33, PID2019-105510GB-C31, PID2019-107847RB-C41, PID2019-107847RB-C42, PID2019-107847RB-C44, PID2019-107988GB-C22, PID2022-136828NB-C41, PID2022-137810NB-C22, PID2022-138172NB-C41, PID2022-138172NB-C42, PID2022-138172NB-C43, PID2022-139117NB-C41, PID2022-139117NB-C42, PID2022-139117NB-C43, PID2022-139117NB-C44 funded by the Spanish MCIN/AEI/ 10.13039/501100011033 and “ERDF A way of making Europe”; the Indian Department of Atomic Energy; the Japanese ICRR, the University of Tokyo, JSPS, and MEXT; the Bulgarian Ministry of Education and Science, National RI Roadmap Project DO1-400/18.12.2020 and the Academy of Finland grant nr. 320045 is gratefully acknowledged. This work was also been supported by Centros de Excelencia ``Severo Ochoa'' y Unidades ``Mar\'{\i}a de Maeztu'' program of the Spanish MCIN/AEI/ 10.13039/501100011033 (CEX2019-000920-S, CEX2019-000918-M, CEX2021-001131-S) and by the CERCA institution and grants 2021SGR00426 and 2021SGR00773 of the Generalitat de Catalunya; by the Croatian Science Foundation (HrZZ) Project IP-2022-10-4595 and the University of Rijeka Project uniri-prirod-18-48; by the Deutsche Forschungsgemeinschaft (SFB1491) and by the Lamarr-Institute for Machine Learning and Artificial Intelligence; by the Polish Ministry Of Education and Science grant No. 2021/WK/08; and by the Brazilian MCTIC, CNPq and FAPERJ.

      The \textit{Fermi}-LAT Collaboration acknowledges support for LAT development, operation and data analysis from NASA and DOE (United States), CEA/Irfu and IN2P3/CNRS (France), ASI and INFN (Italy), MEXT, KEK, and JAXA (Japan), and the K.A.~Wallenberg Foundation, the Swedish Research Council and the National Space Board (Sweden). Science analysis support in the operations phase from INAF (Italy) and CNES (France) is also gratefully acknowledged. This work performed in part under DOE Contract DE-AC02-76SF00515.
      
      We acknowledge the use of public data from the \textit{Swift} data archive.
      
      This work has made use of data from the Joan Oró Telescope (TJO) of the Montsec Observatory (OdM), which is owned by the Catalan Government and operated by the Institute for Space Studies of Catalonia (IEEC).
      
      This research has made use of data from the OVRO 40-m monitoring program \citep{2011ApJS..194...29R}, supported by private funding from the California Institute of Technology and the Max Planck Institute for Radio Astronomy, and by NASA grants NNX08AW31G, NNX11A043G, and NNX14AQ89G and NSF grants AST-0808050 and AST-1109911. 
      S.K. was funded by the European Union ERC-2022-STG - BOOTES - 101076343. Views and opinions expressed are however those of the author(s) only and do not necessarily reflect those of the European Union or the European Research Council Executive Agency. Neither the European Union nor the granting authority can be held responsible for them.

      This work is based on observations with the 100-m telescope of the MPIfR (Max-Planck-Institut für Radioastronomie) at Effelsberg.
      \\ \\
      
      \textit{Author Contributions}. 
      D. Cerasole: project leadership, \textit{Fermi}-LAT data analysis, \textit{Swift} data analysis, multi-wavelength data analysis, theoretical modeling and interpretation, paper drafting and editing; 
      L. Di Venere: supervising the \textit{Fermi}-LAT analysis, paper editing; 
      T. Hovatta: OVRO data analysis;
      E. Lindfors: optical data analysis and collection, theoretical interpretation, paper drafting and editing; 
      S. Loporchio: MAGIC data analysis, theoretical interpretation, paper drafting and editing;
      S. Kiehlmann: OVRO data analysis;
      P. Kouch: optical data analysis and collection;
      L. Pavletic: MAGIC analysis cross-check;
      A. C. S. Readhead: OVRO data analysis.
      \\
      The rest of the authors have contributed in one or several of the following ways: design, construction, maintenance and operation of the instrument(s); preparation and/or evaluation of the observation proposals; data acquisition, processing, calibration and/or reduction; production of analysis tools and/or related Monte Carlo simulations; discussion and approval of the contents of the draft.
\end{acknowledgements}

\bibliographystyle{aa}
\bibliography{bibliography}

\appendix

\section{Details of the intra-band correlation analysis including statistical uncertainties}
\label{sec:appendix_intraband}

In this section, we report details of the intra-band correlation analysis between the flux $F$ and photon index $\Gamma$ in the X-ray and HE $\gamma$-ray bands, presented in Section \ref{subsec:intra-band_correlations}.

Consider a light-curve of a source in a given energy range. Let this be constituted of $n$ temporal bins, $\vec{F}=(F_1, \ldots, F_n)$ and $\vec{\Gamma}=(\Gamma_1, \ldots, \Gamma_n)$ being the arrays of the flux $F$ and photon index $\Gamma$ in the $n$ bins. 
Due to the measurement uncertainties, each $F_{i}$ ($\Gamma_{i}$) is a random variable whose mean value $\bar{F}_i$ ($\bar{\Gamma}_{i}$) and standard deviation $\bar{\sigma}_{F_i}$ ($\bar{\sigma}_{\Gamma_i}$) have been estimated from the data. $\vec{\bar{F}}$ and $\vec{\bar{\sigma}}_{F}$ ($\vec{\bar{\Gamma}}$ and $\vec{\bar{\sigma}}_{\Gamma}$) denote the arrays of the mean values and standard deviations of $F$ ($\Gamma$).
Let $\vec{x}$ be defined as $\vec{x}=\left( \vec{F}, \vec{\Gamma} \right)$, with analogous definitions for $\vec{\bar{x}}$ and $\vec{\bar{\sigma}}_x$.
Since the Pearson coefficient $r$ of the correlation between $F$ (or $\log_{10}F$) and $\Gamma$ is a function of random variables, i.e. $r = r\left( \vec{x} \right)$, $r$ is a random variable itself. The probability density function of $r$ depends on the one of $\vec{x}$ and on the functional form of $r$.
If the function $r\left( \vec{x} \right)$ can be well approximated by its first-order Taylor expansion in a region of size $\vec{\bar{\sigma}}_{x}$ around the mean $\vec{\bar{x}}$, then the mean value $\bar{r}$ of the distribution of $r$ can be approximated by $r_{0} \equiv r\left( \vec{\bar{x}} \right)$, i.e. the Pearson coefficient $r$ computed using the mean values $\vec{\bar{x}}$ \citep{1997sda..book.....C}.
If the uncertainties $\vec{\sigma}_{F}$ and $\vec{\sigma}_{\Gamma}$ are significantly lower than $\vec{\bar{F}}$ and $\vec{\bar{\Gamma}}$, up to factors of a few per cent, this condition is satisfied.
Conversely, when the statistical uncertainties are not negligible and the level of (anti-)correlation indicated by $r_{0}$ is non-zero, the functional form of the Pearson coefficient shows significant non-linearity in a region around $\vec{\bar{x}}$ of a size comparable to $\vec{\bar{\sigma}}_{x}$, thus preventing us to employ $r_{0}$ to approximate $\bar{r}$ safely. 
In this case, statistical uncertainties lower the level of (anti-)correlation compared to $r_{0}$ and Monte-Carlo simulations have to be used to estimate $\bar{r}$ properly \citep{1997sda..book.....C}.
An example of this effect can be shown using the scatter plots of the photon index and flux at HE $\gamma$ rays in Figure \ref{fig:Fermi_LAT_correlations}.
The best-fit values $\vec{\bar{F}}$ and $\vec{\bar{\Gamma}}$ would indicate high levels of correlation in the `Pre-flare', `Flare' and `Post-flare' periods, the $r_0$ values between $\log_{10} F$ and $\Gamma$ being equal to $0.89$, $0.71$ and $0.93$, respectively.
However, if the statistical uncertainties around $\vec{\bar{F}}$ ($\vec{\bar{\Gamma}}$) are included by extracting flux (index) values from distributions centered at $\vec{\bar{F}}$ ($\vec{\bar{\Gamma}}$) and with standard deviation $\vec{\bar{\sigma}}_{F}$ ($\vec{\bar{\sigma}}_{\Gamma}$), the simulated flux-index pairs tend to show, on average, higher scattering around the general trend compared to $\vec{\bar{F}}$ and $\vec{\bar{\Gamma}}$, therefore indicating a lower correlation compared to $r_{0}$.

In the simulations, we assumed that the joint probability density function of the fluxes $\vec{F}$ and indices $\vec{\Gamma}$ can be approximated by independent Gaussian functions centered at $\vec{\bar{F}}$ and $\vec{\bar{\Gamma}}$, respectively, with standard deviations given by $\vec{\bar{\sigma}}_{F}$ and $    \bar{\vec{\sigma}}_{\Gamma}$.
Each light-curve was simulated by extracting the $\vec{F}$ and $\vec{\Gamma}$ arrays from the assumed joint probability density function and the corresponding Pearson coefficient $r=r\left( \log_{10} \vec{F}, \vec{\Gamma} \right)$ was computed.
A large number $N$ of light-curves was simulated and these were employed to estimate $\bar{r}$. The guidelines for choosing $N$ are reported below.
In the case of the correlation between $\log_{10} F_{\gamma}$ and $\Gamma_{\gamma}$ at HE $\gamma$ rays in Figure \ref{fig:Fermi_LAT_correlations}, the $\bar{r}$ values in the `Pre-flare', `Flare' and `Post-flare' periods were found to be equal to $0.41$, $0.32$ and $0.25$ (Table \ref{tab:summ_table_Fermi_correlations}), respectively, thus indicating significantly lower correlation than the corresponding $r_0$ values reported above. We checked that assuming log-normal distributions with mode $\vec{\bar{F}}$ and $\vec{\bar{\sigma}}_{F}$ rather than Gaussian functions for the flux $\vec{F}$ leads to estimates of $\bar{r}$ compatible within a few per cent with those in Table \ref{tab:summ_table_Fermi_correlations}.
Conversely, in the case of the correlation between $\log_{10} F_{X}$ and $\Gamma_{X}$ in the X-ray band (Figure \ref{fig:harder-when-brighter}), for each period the differences between $\bar{r}$ and $r_0$ were found to be below 0.01, due to the low relative uncertainties in X rays (Table \ref{tab:results_XRT_B2}).

To evaluate the significance of the correlation quantified by $\bar{r}$ under the null hypothesis of non-correlation, we considered the $N$ light-curves simulated as described above and, for each light-curve, the $\vec{F}$ and $\vec{\Gamma}$ arrays were shuffled randomly and independently. In this way, we obtained a set of $N$ light-curves in which the flux and index are uncorrelated. We used these to construct the distribution of Pearson coefficients under the null hypothesis, from which we derived the $p$-value and significance of $\bar{r}$.
To probe a correlation with a given $p$-value, we used a number of simulations $N\gtrsim100/p$-value. 
It is worth mentioning that including the statistical uncertainties has no major effect on the distributions of the Pearson coefficient under the hypothesis of non-correlation.

\section{Comparison between the one-zone and two-zone model solutions}
\label{sec:appendix_chi_square}

The estimation of the parameters of the one-zone and two-zone model solutions presented in Figures \ref{fig:B2_modeling_one_zone} and \ref{fig:B2_modeling_two_zones}, respectively, was performed using a `fit-by-eye' strategy.
Nonetheless, we report here quantitative indications that the two-zone model is more adequate to describe the broad-band SED than the one-zone model.
Table \ref{tab:chi_square} reports the $\chi^2$ values for the one-zone and two-zone solutions, computed using the spectral points from \textit{Swift}-UVOT, \textit{Swift}-XRT, \textit{Fermi}-LAT and MAGIC.
The radio band was not included in Table \ref{tab:chi_square} since the one-zone model does not reproduce the spectrum at the lowest frequencies, as discussed in Section \ref{sec:modeling}.
The $\chi^2$ values computed in the optical/UV band and in X rays are higher than those in $\gamma$ rays. This is a consequence of the small uncertainties of the spectral points in the optical/UV band and in X rays with respect to those at HE and VHE $\gamma$ rays. 
The $\chi^2$ values computed for the two-zone solution are lower than those obtained for the one-zone model, indicating that the former provides a better agreement with the data.
However, a more sophisticated procedure should be employed to statistically claim which model provides a more accurate description of the MWL data. 
Such a procedure should take into account the number of free parameters involved in the optimization and the systematic uncertainties on the measurements from each instrument.
Given the high dimensionality of the model parameter space, such a procedure is outside the scope of this work.

\begin{table} [ht]
\small
\caption{$\chi^2$ comparison of the one-zone and two-zone SSC model solutions.}
\label{tab:chi_square}
\begin{center}
\begin{tabular}{ccc}
\hline
\multicolumn{1}{c}{} &
\multicolumn{1}{c}{$\chi^2$ (One-zone model)} &
\multicolumn{1}{c}{$\chi^2$ (Two-zone model)} \\
\hline
\textit{Swift}-UVOT & 245 & 176 \\
\textit{Swift}-XRT & 163 & 51 \\
\textit{Fermi}-LAT & 0.32 & 0.049 \\
MAGIC & 1.76 & 0.359 \\
\hline
\end{tabular}
\end{center}
\tablefoot{The $\chi^2$ values are reported separately for the \textit{Swift}-UVOT, \textit{Swift}-XRT, \textit{Fermi}-LAT and MAGIC spectral points.}
\end{table}

\section{Details of the multi-wavelength analysis}

In this section, we provide more detail on the observations and the analyses performed in the different energy ranges.

\begin{table*}[ht]
\caption{Summary of MAGIC observations of \btwo.}
\small
\label{tab:magic_table}
\begin{center}
\begin{tabular}{ccccccc}
\hline
\multicolumn{1}{c}{Start time} &
\multicolumn{1}{c}{Stop time} &
\multicolumn{1}{c}{Observing time} &
\multicolumn{1}{c}{Zenith range} &
\multicolumn{1}{c}{Significance} &
\multicolumn{1}{c}{Flux ($E > 135\,\text{GeV}$)} &
UL \\
\multicolumn{1}{c}{[MJD]} &
\multicolumn{1}{c}{[MJD]} &
\multicolumn{1}{c}{[h]} &
\multicolumn{1}{c}{[deg]} &
\multicolumn{1}{c}{} &
\multicolumn{1}{c}{[10$^{-11}$ ph cm$^{-2}$ s$^{-1}$]} &
\multicolumn{1}{c}{[10$^{-11}$ ph cm$^{-2}$ s$^{-1}$]}\\
\hline
59126.84 & 59126.86 & 0.48 & $20-30$ & $4.7\sigma$ & 3.55 $\pm$ 1.50 & - \\
59127.84 & 59127.88 & 0.73 & $20-37$ & $1.2\sigma$ & - & 3.37\\
59130.93 & 59130.97 & 0.98 & $50-64$ & $-0.7\sigma$ & - & 14.1\\
59131.83 & 59131.95 & 1.96 & $20-64$ & $4.3\sigma$ & 3.32 $\pm$ 1.15 & - \\
59132.86 & 59132.91 & 1.30 & $30-50$ & $2.1\sigma$ & - & 6.17\\
\hline
\end{tabular}
\end{center}
\tablefoot{For each night, the observing time after quality cuts, zenith range, significance of the excess of $\gamma$-ray signal from the source and gamma-ray flux integrated above 135 GeV are reported. Upper limits (ULs) at 95\% confidence level are provided for the observations with significance of the $\gamma$-ray signal from the source below $3\sigma$.}
\end{table*}

\begin{table*} [ht]
\small
\caption{Summary of the curvature tests performed on the HE $\gamma$-ray spectra from \textit{Fermi}-LAT data.}
\label{tab:summ_table_spectral_B2_LP_PLEC_BPL}
\begin{center}
\begin{tabular}{cccccccc}
\hline
\multicolumn{1}{c}{Period} &
\multicolumn{1}{c}{Start} &
\multicolumn{1}{c}{Stop} &
\multicolumn{1}{c}{$\text{TS}_{\text{curv, LP}}$} &
\multicolumn{1}{c}{$\text{TS}_{\text{curv, PLEC}}$} & 
\multicolumn{1}{c}{$\text{TS}_{\text{curv, BLP}}$} \\
\multicolumn{1}{c}{}&\multicolumn{1}{c}{[MJD]}&\multicolumn{1}{c}{[MJD]} & \multicolumn{1}{c}{} & \multicolumn{1}{c}{} & \multicolumn{1}{c}{} \\
\hline
Pre-flare & 54682 & 58940 & 4.1 & 6.6 & 2.0 \\
Flare & 58940 & 59190 & 4.8 & 7.7 & 1.7 \\
Post-flare & 59190 & 59945 & 2.3 & 2.2 & 1.5 \\
\hline
\end{tabular}
\end{center}
\end{table*}

\begin{table*}
\caption{
Results of the correlation analysis between the spectral index and flux in the HE $\gamma$-ray band.} 
\small
\label{tab:summ_table_Fermi_correlations}
\begin{center}
\begin{tabular}{cccccccc}
\hline
\multicolumn{1}{c}{Light-curve} & 
\multicolumn{1}{c}{Period} & 
\multicolumn{1}{c}{$r$} & 
\multicolumn{1}{c}{$p$-value} & 
\multicolumn{1}{c}{Significance} & 
\multicolumn{1}{c}{$p_{0}$} & 
\multicolumn{1}{c}{$p_{1}$} & 
\multicolumn{1}{c}{$\chi^{2}/\text{ndof}$} \\
\hline
30-day bins & Pre-flare & $0.41$ & $0.033$ & $2.1\sigma$ &  $7 \pm 2$ & $0.7 \pm 0.2$ & $4.1/20$ \\
Daily bins & Flare & $0.32$ & $0.10$ & $1.7\sigma$ & $7 \pm 2$ & $0.8 \pm 0.3$ & $6.4/16$ \\
30-day bins & Post-flare & $0.25$ & $0.32$ & $1.0\sigma$ & $8 \pm 5$ & $0.9 \pm 0.7$ & $0.3/4$ \\
\end{tabular}
\end{center}
\tablefoot{Columns:
(1) $\textit{Fermi}$-LAT light-curve.
(2) Period. (3) Pearson coefficient between $\Gamma_{\gamma}$ and $\log_{10} F_\gamma$, estimated as described in Appendix \ref{sec:appendix_intraband}. (4), (5) $p$-value and significance for the null hypothesis of no-correlation. (6), (7) Best-fit parameters of the $\Gamma_{\gamma}-\log_{10} F_\gamma$ trend with Eq. \ref{eq:X}. (8) Goodness of fit, expressed as $\chi^2$ per number of degrees of freedom (ndof).}
\end{table*}

\begin{table*}
\small
\caption{Results of the VHE $\gamma$-ray spectral analyses of MAGIC data.}
\label{tab:summ_table_spectral_MAGIC_B2_stateA_B}
\begin{center}
\begin{tabular}{cccccccc}
\hline
\multicolumn{1}{c}{Period} &
\multicolumn{1}{c}{Start time} &
\multicolumn{1}{c}{Stop time} &
\multicolumn{1}{c}{$\Gamma_{\text{PL}}$} &
\multicolumn{1}{c}{N$_0\,\times\,10^{-10}$} &
\multicolumn{1}{c}{$E_0$}\\
\multicolumn{1}{c}{}&\multicolumn{1}{c}{[MJD]}&\multicolumn{1}{c}{[MJD]} & \multicolumn{1}{c}{} & \multicolumn{1}{c}{[$\text{TeV}^{-1}\, \text{cm}^{-2}\, \text{s}^{-1}$]} & [GeV]\\
\hline
A & 59126.84 & 59127.88  & $4.16\pm0.63$ & $4.21\pm1.51$ & 130.95 \\
B & 59130.93 & 59132.91 & $3.75\pm0.40$ & $7.36\pm1.99$ & 125.16 \\
\hline
\end{tabular}
\end{center}
\end{table*}

\begin{table*}
\small
\caption{Results of the HE $\gamma$-ray spectral analyses of \textit{Fermi}-LAT data on \btwo\ in Periods A and B.} 
\label{tab:summ_table_spectral_B2_stateA_B}
\begin{center}
\begin{tabular}{cccccccc}
\hline
\multicolumn{1}{c}{Period} &
\multicolumn{1}{c}{Start time} &
\multicolumn{1}{c}{Stop time} &
\multicolumn{1}{c}{$\Gamma_{\text{PL}}$} &
\multicolumn{1}{c}{F ($E>\,100\,\text{MeV}$)$\,\times\,$10$^{-8}$ } &
\multicolumn{1}{c}{TS$_{\text{PL}}$}\\
\multicolumn{1}{c}{}&\multicolumn{1}{c}{[MJD]}&\multicolumn{1}{c}{[MJD]} & \multicolumn{1}{c}{} & \multicolumn{1}{c}{[ph cm$^{-2}$ s$^{-1}$]} \\
\hline
A & 59126.39 & 59128.39 & $2.5\pm0.3$ & $21\pm7$ & 24 \\
B & 59130.39 & 59133.39 & $1.7\pm0.2$ & $8\pm4$ & 37 \\
\hline
\end{tabular}
\end{center}
\end{table*}

\begin{table*}
\small
\caption{Results from the spectral analyses of the \textit{Swift}-XRT data from the \btwo\ high state.}
\label{tab:results_XRT_B2}
\begin{center}
\begin{tabular}{cccccccc}
\hline
\multicolumn{1}{c}{Start time} &
\multicolumn{1}{c}{Exposure} &
\multicolumn{1}{c}{$C_{\text{PL}}/\text{dof}$} &
\multicolumn{1}{c}{$C_{\text{BPL}}/\text{dof}$} &
\multicolumn{1}{c}{$F$-statistic/$p$-value} &
\multicolumn{1}{c}{$\Gamma_{\text{PL}}$ ($\Gamma_{1, \text{BPL}}, \Gamma_{2, \text{BPL}}$)} &
\multicolumn{1}{c}{$E_{\text{break}}$} &
\multicolumn{1}{c}{Flux [0.3 - 10 keV]} \\
\multicolumn{1}{c}{[MJD]} & \multicolumn{1}{c}{[s]} & \multicolumn{1}{c}{} & \multicolumn{1}{c}{} & \multicolumn{1}{c}{} & \multicolumn{1}{c}{} & \multicolumn{1}{c}{[keV]} & \multicolumn{1}{c}{[10$^{-12}$ erg cm$^{-2}$ s$^{-1}$]} \\
\hline
59125.442 & 2476 & 221/245 & 220/243 & 0.55/0.58 & $2.54\pm0.06$ &      -      & $21.3\pm0.8$ \\
59128.283 & 1303 & 125/153 & 124/151 & 0.61/0.54 & $2.50\pm0.11$ &      -      & $11.0\pm0.6$ \\
59132.286 & 1171 & 126/149 & 126/147 & 0.58/0.56 & $2.58\pm0.10$ &      -      & $11.4\pm0.6$ \\
59134.853 & 1489 & 171/201 & 171/199 & 0.58/0.56 & $2.56\pm0.08$ &      -      & $23.5\pm1.1$ \\
59136.848 & 2003 & 246/271 & 235/269 & 6.30/2.1$\times10^{-3}$ & $2.09\pm0.09$,
                                                $2.83\pm0.26$ & $1.9\pm0.4$ & $24.5\pm2.7$ \\
59138.851 & 1841 & 253/271 & 237/269 & 9.08/1.5$\times10^{-4}$ & $2.14\pm0.08$,
                                                $4.72\pm1.13$ & $3.0\pm0.4$ & $18.4\pm3.5$ \\
59156.828 & 1948 & 277/322 & 276/320 & 0.57/0.56 & $2.22\pm0.04$ &      -      & $73.0\pm2.1$ \\
59157.823 & 1839 & 242/248 & 241/246 & 0.51/0.60 & $2.36\pm0.06$ &      -      & $26.4\pm1.0$ \\

\hline
\end{tabular}
\end{center}
\tablefoot{Cash statistic values assuming PL and BPL models for the source intrinsic spectrum are reported, as well as the corresponding $F$-statistic values and $p$-values for the spectral curvature test.
The BPL model was preferred over the single PL model if $p\text{-value}\,<2.7\times10^{-3}$, corresponding to $3\sigma$ confidence level.
For the observations with BPL model favored over the PL one, the BPL best-fit spectral indices $\Gamma_{1, \text{BPL}}, \, \Gamma_{2, \text{BPL}}$ and energy break $E_{\text{break}}$ are indicated rather than the PL index $\Gamma_{\text{PL}}$.
The normalization energy $E_{0}$ of the PL and BPL functional forms in Equations \ref{eq:PL} and \ref{eq:BPL} was set to $E_{0} = 1\,\text{keV}$.}
\end{table*}

\begin{table*}
\small
\caption{Intrinsic magnitudes from the spectral analyses on \btwo\ \textit{Swift}-UVOT observations during the 2020 $\gamma$-ray outburst.}
\label{tab:summ_table_UVOT_obs_B2}
\begin{center}
\begin{tabular}{cccccccc}
\hline
\multicolumn{1}{c}{Start time} & \multicolumn{1}{c}{Exposure} &
\multicolumn{1}{c}{V}&\multicolumn{1}{c}{B}&\multicolumn{1}{c}{U}&\multicolumn{1}{c}{W1}&\multicolumn{1}{c}{M2}&\multicolumn{1}{c}{W2}\\
\multicolumn{1}{c}{[MJD]}&\multicolumn{1}{c}{[s]}&\multicolumn{1}{c}{[mag]}&\multicolumn{1}{c}{[mag]}&\multicolumn{1}{c}{[mag]}&\multicolumn{1}{c}{[mag]}&\multicolumn{1}{c}{[mag]}&\multicolumn{1}{c}{[mag]}\\
\hline
59125.442 & 2473 & - & - & - & - & - & $14.60\pm0.02$ \\
59128.283 & 1260 & - & - & $14.83\pm0.02$ & - & - & -\\ 
59132.286 & 1168 & - & - & $14.69\pm0.02$ & - & - & -\\
59134.853 & 1449 & $15.34\pm0.05$ & $15.62\pm0.03$ & $14.69\pm0.03$ & $14.46\pm0.03$ & $14.29\pm0.03$ & $14.48\pm0.03$\\ 
59136.848 & 1953 & $15.34\pm0.04$ & $15.64\pm0.03$ & $14.63\pm0.03$ & $14.46\pm0.03$ & $14.33\pm0.03$ & $14.48\pm0.02$\\ 
59138.851 & 1787 & $15.44\pm0.04$ & $15.67\pm0.03$ & $14.74\pm0.03$ & $14.58\pm0.03$ & $14.43\pm0.03$ & $14.57\pm0.03$\\ 
59156.828 & 2003 & $15.54\pm0.06$ & $15.78\pm0.03$ & $14.85\pm0.03$ & $14.64\pm0.03$ & $14.50\pm0.03$ & $14.66\pm0.03$\\ 
59157.823 & 1798 & $15.56\pm0.06$ & $15.87\pm0.03$ & $14.92\pm0.03$ & $14.72\pm0.03$ & $14.66\pm0.03$ & $14.78\pm0.03$\\ 
\hline
\end{tabular}
\end{center}
\end{table*}

\end{document}